\documentclass[11pt,a4paper]{article}
\usepackage{amsmath,amssymb,amsfonts,amsthm}
\usepackage[hidelinks,breaklinks=true]{hyperref}
\usepackage{authblk}
\usepackage[english]{babel}
\usepackage[utf8]{inputenc}
\usepackage{xcolor}
\usepackage{caption}
\usepackage{sectsty}
\usepackage{cite}
\usepackage{subcaption}
\usepackage{graphicx}
\usepackage{dcolumn}
\usepackage{bm}
\usepackage[font=small]{caption}

\subsectionfont{\normalfont\bf}
\subsubsectionfont{\normalfont\itshape}

\newcommand{\R}{\mathbb{R}}
\newcommand{\Ax}{\mathcal{A}}

\newcommand{\Ho}{\mathcal{H}}

\newcommand{\bi}{\begin{itemize}}
\newcommand{\ei}{\end{itemize}}
\newcommand{\be}{\begin{equation}}
\newcommand{\ee}{\end{equation}}
\newcommand{\rhomax}{\rho_{\rm MAX}}
\newcommand{\bsigma}{{\bar\sigma}}
\newcommand{\bomega}{{\bar\omega}}

\usepackage{caption}
\usepackage{comment}

\begin{document}
\title{\bf\LARGE Periodic analogues of the Kerr solutions:\\ a numerical study}

\author[1]{Javier Peraza \thanks{jperaza@cmat.edu.uy}}
\author[1]{Martín Reiris \thanks{mreiris@cmat.edu.uy}}
\affil{Universidad de la República, Uruguay
}
\author[2]{Omar E. Ortiz \thanks{omar.ortiz@unc.edu.ar}}
\affil{FAMAF, Universidad Nacional de Córdoba, Argentina}
\date{}

\renewcommand\Affilfont{\itshape\small}
\maketitle
\begin{abstract}
In recent years black hole configurations with non standard topology or with
non-standard asymptotic have gained considerable attention. In this article we
carry out numerical investigations aimed to find periodic coaxial configurations
of co-rotating 3+1 vacuum black holes, for which existence and uniqueness has
not yet been theoretically proven. The aimed configurations would extend
Myers/Korotkin-Nicolai's family of non-rotating (static) coaxial arrays of black
holes. We find that numerical solutions with a given value for the area $A$ and
for the angular momentum $J$ of the horizons appear to exist only when the
separation between consecutive horizons is larger than a certain critical value
that depends only on $A$ and $|J|$. We also establish that the solutions have
the same Lewis's cylindrical asymptotic as Stockum's infinite rotating cylinders.
Below the mentioned critical value the rotational energy appears to be too big
to sustain a global equilibrium and a singularity shows up at a finite distance
from the bulk. This phenomenon is a relative of Stockum's asymptotic's
collapse, manifesting when the angular momentum (per unit of axial length)
reaches a critical value compared to the mass (per unit of axial length), and
that results from a transition in the Lewis's class of the cylindrical exterior
solution. This remarkable phenomenon seems to be unexplored in the context of
coaxial arrays of black holes. Ergospheres and other global properties are also
presented in detail. 
\end{abstract}
\section{Introduction}
In recent years vacuum black hole configurations with non standard topology or
with non-standard asymptotic have gained considerable attention. In five dimensions, Emparan and Reall \cite{Emparan:2001wn} have found
asymptotically flat stationary solutions with ring-like $S^{1}\times S^{2}$
horizon and Elvang and Figueras \cite{Elvang:2007rd} have found asymptotically flat black Saturns, where a black ring $S^1 \times S^2$ rotates around a black sphere $S^3$. More recently Khuri, Weinstein and Yamada \cite{Khuri:2020dbw, Khuri:2021fqu},
have found periodic static coaxial arrays of horizons either with spherical
$S^{3}$ or ring-like $S^{1}\times S^{2}$ topology. Rather than
asymptotically flat, these latter configurations have Levi-Civita/Kasner
asymptotic and generalize to five dimensions the important
Myers/Korotkin-Nicolai (MKN) family of vacuum static $3+1$ solutions \cite{Myers, KN, Korotkin:1994cp},
referred sometimes as ``periodic Schwarzschild'', as their are obtained as ``linear" superpositions of Schwarzschild's solutions via Weyl's method. The search for new solutions has branched rapidly to higher dimensions and to different fields, giving by now a rich landscape of topologies, \cite{Emparan:2008eg}. 
However, the question whether ``periodic Kerr'' configurations generalizing the ``periodic Schwarzschild'' exist, either in four or higher dimensions, remains still open. In this article we aim to investigate this latter problem in four dimensions numerically. Specifically, we carry out numerical investigations pointing at constructing periodic coaxial configurations of co-rotating $3+1$ black holes.
\begin{figure}[ht]
\centering
\captionsetup{margin=2cm}
\includegraphics[scale=.2]{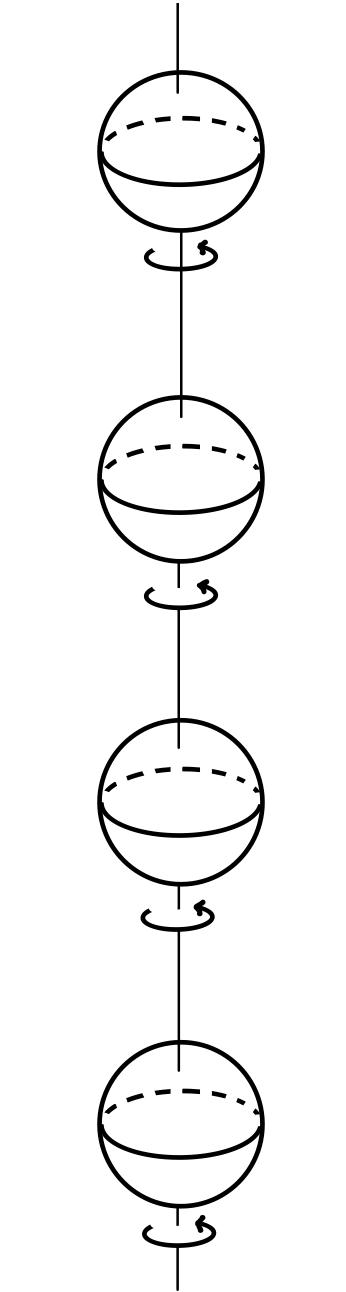}
\caption{Periodic configuration of co-rotating coaxial black holes. The horizons are equally spaced, have the same area $A$ and angular momentum $J$.}
\label{fig1}
\end{figure}
The equally spaced horizons are all intended to have the same area $A$, angular momentum $J$ and angular velocity $\Omega_{H}$, (see Figure \ref{fig1}). This problem has been discussed by Korotkin and Nicolai in \cite{Korotkin:1994cp} (section 4) and also by Mavrin in \cite{Mav}, however non conclusively. Numerical investigations and an analysis of the solution space and of the global properties of the solutions appear to be carried out here for the first time. 

This problem poses a number of difficulties on the numerical side. The central, relevant equation to solve is an harmonic map in two dimensions, whose solutions are found as the stationary regime of a heat-type  flow for the harmonic map. The numerical study of the problem allowed us to find the correct formulation of the asymptotic boundary conditions, which constituted the main difficulty of the problem.
\vspace{.2cm}

The results of this paper can be summarized as follows. Periodic
configurations having a given value of $J$ and $A(>8\pi |J|)$, appear to exist
only when the separation between the horizons is larger than a certain critical
value that depends only on $A$ and $|J|$. The smaller the
separation the bigger the angular velocity $|\Omega_{H}|$ and the bigger the
rotational energy and the total energy. The
asymptotic (suitably defined) matches always a
Lewis's solution \cite{Lewis} as in the Stockum's rotating cylinders \cite{vanStockum:1937zz}.
Furthermore, as the separation between horizons gets smaller than the critical value, no
asymptotic can hold the given amount of energy and we evidence a singularity
formation at a finite distance from the bulk. Roughly speaking, the asymptotic
``closes up'' due to too much rotational energy. This phenomenon is similar to
Stockum's asymptotic collapse, studied in \cite{vanStockum:1937zz}, and manifesting when the angular momentum (per
unit of axial length) increases past a critical value compared to the mass (per unit of
axial length). When the rotation increases, the exterior Lewis solution transits from one extending to infinity to one blowing up at a finite distance from the material cylinder. There is a change in the Lewis's class. This change of class, that prevents black holes from
getting too close, appears to be entirely novel in our context and is in sharp contrast to what
occurs for the periodic Schwarzschild configurations, where horizons can get arbitrarily
near. The ergo-regions are always bounded and disjoint. We observe though that below the critical separation the ergospheres can indeed merge, but such solutions do not extend to infinity. The Komar mass $M$ per black hole satisfies the relevant inequality,
\begin{equation}
\sqrt{\frac{A}{16\pi}+\frac{4\pi J^{2}}{A}}\leq M,
\end{equation}
and equality is approached as the separation between the black holes grows unboundedly and the geometry near the horizons
approaches that of Kerr.
\vspace{.2cm}

The paper is organized as follows. In section \ref{background} we overview the theoretical and numerical problem commenting on the difficulties and strategies. We present the main equations to be solved and recall the Lewis's classes. We also discuss the boundary condition for the harmonic map heat flow and give the precise set up for the numerical study. In section \ref{sec_numerical_implementation} we discuss the numerical techniques used to solve the equations. In section \ref{sec_results} we present our results, discussing several features of the numerical solutions obtained, and in section \ref{conclusions} we contextualize our work and give some possible future directions to our study.

\section{The setup}\label{background}
\subsection{Overview of the mathematical and numerical problem}
The configurations that we are looking for have three degrees of freedom that we will take to be the area $A$, the angular momentum $J$, and the period $L$, with $L$ linked to the physical separation between consecutive black holes. These parameters need to be incorporated into the boundary conditions of the main equations. The stationary and axisymmetric metrics are assumed to be in Weyl-Papapetrou form and are recalled in (\ref{asm}), (see for instance \cite{Wei90}). In this expression $\rho\in \mathbb{R}^{+}$ and $z\in \mathbb{R}$ are the Weyl-Papapetrou cylindrical coordinates and play a role in what follow. For Weyl-Papapetrou metrics, the Einstein equations reduce to find an axisymmetric harmonic map $(\omega(\rho, z),\eta(\rho,z)):\mathbb{R}^{3}\rightarrow \mathbb{H}^{2}$ from $\mathbb{R}^{3}$ into the hyperbolic space $\mathbb{H}^{2}$. Here $\eta$ is the norm squared of the axisymmetric Killing field and $\omega$ is the twist potential. All the metric components can be recovered from $\eta$ and $\omega$ after line-integrations. The full set of reduced equations are presented in section \ref{RSE}. The harmonic map equations are (\ref{REE10})-(\ref{REE20}) and the metric components are (\ref{WV}) after the line integrations of (\ref{quadOmega}) and (\ref{quadgamma}).  

\begin{figure}[h]
\centering
\captionsetup{margin=2cm}
\includegraphics[scale=0.25]{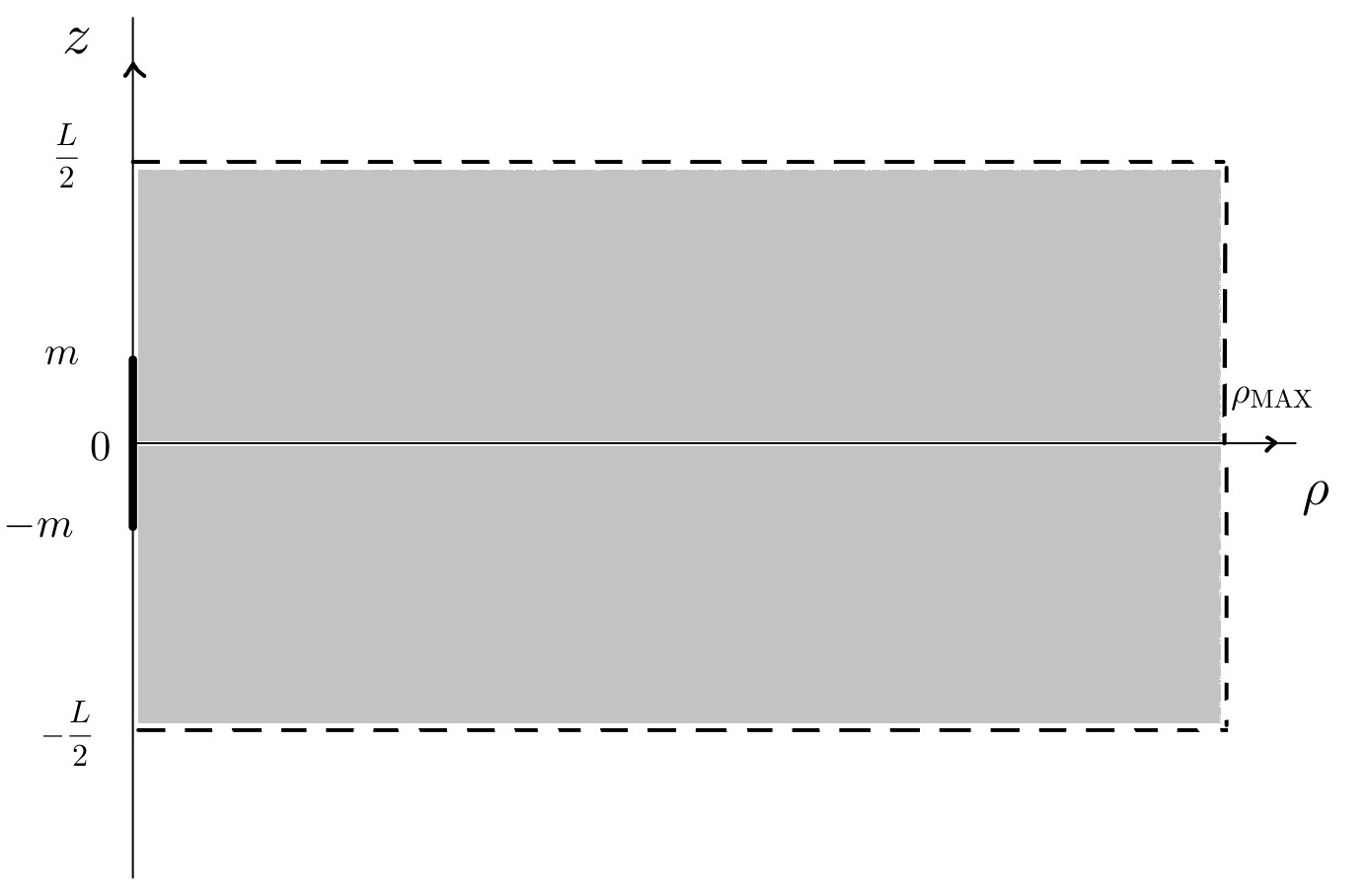}
\caption{The numerical domain with the horizon marked as a thick interval.}
\label{domain}
\end{figure}

As we look for metrics periodic in $z$, we restrict the analysis to the region $\{\rho\geq 0, -L/2\leq z\leq L/2\}$ keeping appropriate periodicity conditions on the top and bottom lines $\{\rho\geq 0,\ z=\pm L/2\}$. Here $L$ is the period mentioned earlier and is a free parameter. The harmonic map equations for $\omega$ and $\eta$ need to be supplied with boundary data on the border $\{\rho=0, -L/2\leq z \leq L/2\}$. This boundary contains the horizon $\mathcal{H}=\{\rho=0,\ -m\leq z\leq m\}$, and the two components of the axis ${\mathcal{A}}$ which is the complement of $\mathcal{H}$. The boundary conditions arise from two sides: from the natural conditions that $\eta$ and $\omega$ must verify on horizons and axes, and from fixing the values of the area $A$ and of the angular momentum $J$. The angular momentum $J$ is set by specifying Dirichlet data for $\omega$ on the axis, whereas the area $A$ is set by specifying the limit values of $\eta/\rho^{2}$ at the poles $z=\pm m$. The fact that one can incorporate $A$ and $J$ into the boundary conditions in this simple way is a great advantage of the formulation. The parameter $m$ that may seem to be a free parameter, is equal to the surface gravity $\kappa$, ($\kappa^{2}:=-\langle \nabla \partial_{t},\nabla \partial_{t}\rangle/2$), of the horizons times $A/4\pi$ and therefore can be fixed using the freedom in the definition of the stationary Killing field $\partial_{t}$, or the definition of time in (\ref{asm}). In this article we chose $\kappa=\kappa(A,J)$ equal to the surface gravity of the Kerr black holes for the given $A$ and $J$.      

The full set of boundary conditions on $\{\rho=0, -L/2\leq z \leq L/2\}$ is discussed in section \ref{sec_boundary_conditions}. 

In practice, the harmonic map equations need to be solved numerically on a finite rectangle $[0,\rho_{\rm\small{MAX}}]\times [-L/2,L/2]$, and this adds the extra difficulty of finding natural boundary conditions also at $\rho=\rho_{\rm MAX}$, for $\omega$ and for $\eta$. On physical grounds one expects the solutions to become asymptotically independent of $z$ and indeed approaching the Lewis models of Stockum's rotating cylinders that we recall in section \ref{asymptotic_models}. But the problem is that one does not know a priori which Lewis solution shows up for the given $A$, $J$ and $L$. If that information were known, then one could easily supply appropriate boundary conditions at $\rho_{\rm MAX}$. Now, all Lewis solutions have $\omega = wz $, so we make $\omega(\rho_{\rm MAX},z)=wz$. For $\eta$ however not such single choice is possible. We set a Neumann type of boundary condition for $\eta$ and for that we use that, on actual solutions, the Komar mass expression $M(\rho)$, in (\ref{asymptotic_condition_sigma}), is constant, so that $M(\rho_{\rm MAX})=M(\rho)$ easily relates $\partial_{\rho}\eta(\rho_{\rm MAX})$ to $\eta$ and $\omega$ at any $\rho<\rho_{\rm MAX}$. Then, to define the condition for $\eta$ we make use of this relation, equating $\partial_{\rho}\eta(\rho_{\rm MAX})$ to the average of the Komar mass expression in the bulk $0<\rho<\rho_{\rm MAX}$. We discuss this condition in section \ref{sec_boundary_conditions}, and state it in (\ref{asymptotic_condition_sigma}). This peculiar Neumann condition for $\eta$ at $\rho_{\rm MAX}$ is what gave us the best numerical results in terms of the speed and the stability of our code.

To find solutions for the harmonic map system, we run a harmonic map heat flow with certain initial data, and look for the stationary solutions in the long time. We can show that the numerical solutions for two different values of $\rho$, say $1\ll \rho_{\rm MAX}^{1}<\rho_{\rm MAX}^{2}$, but the same values of $A$, $J$ and $L$, overlap on the smaller rectangle $[0,\rho^{1}_{\rm MAX}]\times [-L/2,L/2]$. This shows that the solutions constructed indeed depend on $A$, $J$ and $L$ only and the boundary condition at $\rho_{MAX}$ does not introduce any spurious new degree of freedom, but just drives the harmonic map heat flow to settle with the right asymptotic for the given $A$, $J$ and $L$.  

\subsection{The reduced stationary equations}\label{RSE}

The black hole configurations that we are looking for will be in Weyl-Papapetrou form,
\begin{equation} \label{asm}
g = -V dt^2 + 2 W dt d\phi + \eta d\phi^2 +  \frac{e^{2 \gamma}}{\eta}  (d\rho^2 + dz^2),
\end{equation}
where $(t, \rho ,z , \phi) \in \R \times \mathbb{R}^{+}\times \mathbb{R} \times S^1$ and where the components $V,W,\eta$ and $\gamma$ depend only on $(\rho ,z)$. We require $V, W, \eta$ and $\gamma$ to be $z$-periodic with period $L$, and, to prevent struts on the axis, we demand in addition $V,W,\eta$ and $\gamma$ to be symmetric with respect to the reflection $z\rightarrow -z+ L$.

If we denote by $\omega$ the twist potential of $\partial_{\varphi}$, then $\eta(\rho,z)$ and $\omega(\rho,z)$ satisfy the harmonic map equations (Ernst equations),
\begin{align}
\Delta_\rho \eta &= \frac{|\partial \eta|^2 - |\partial \omega |^2}{\eta},
\label{REE10} \\
\Delta_\rho \omega &=  2 \langle \partial \omega, \frac{\partial \eta}{\eta}\rangle 
\label{REE20},
\end{align}
where the inner product $\langle\ ,\ \rangle$ is with respect to the flat metric $d\rho^2 + dz^2$, and $\Delta_\rho = \partial^2_\rho + \frac{1}{\rho} \partial_\rho + \partial^2_z$ is the $3$-dim Laplacian in cylindrical coordinates $(\rho,z,\phi)$ under axial symmetry. The metric components $W$ and $V$ follow from the identities,
\begin{equation} \label{WV}
W = \eta \Omega, \quad V  = \frac{\rho^2 - W^2}{\eta},
\end{equation} 
after finding the angular velocity function $\Omega$ by line integration of,
\begin{align}\label{quadOmega}
\Omega_z = \rho \frac{\omega_{\rho}}{\eta^2},\quad \Omega_\rho = -\rho
\frac{\omega_z}{\eta^2}.
\end{align}
The exponent $\gamma$ is found after line integration of,
\begin{align}\label{quadgamma}
\gamma_z = \frac{\rho}{2 \eta^2} (\eta_\rho \eta_z + \omega_\rho \omega_z), 
\quad \gamma_\rho = \frac{\rho}{4 \eta^2} (\eta_\rho^2 - \eta_z^2 + \omega_\rho^2 - \omega_z^2).
\end{align}
Thus, the Einstein equations are solved in a ladder-like fashion: first solve the harmonic map equations \eqref{REE10}-\eqref{REE20}, then solve the quadratures \eqref{quadOmega} and \eqref{quadgamma}, and finally get from them the other coefficients $V$ and $W$ of the metric \eqref{asm}.

\subsection{Lewis solutions and the asymptotic models}\label{asymptotic_models}

The Lewis solutions \cite{Lewis} (see also \cite{vanStockum:1937zz}) are cylindrically symmetric (i.e. independent on $\varphi$ and $z$) rotating stationary vacuum solutions. Some of the solutions extend to infinity and some do not. The possible forms for $\eta$ are the following,
\begin{align}
\label{I} {\rm (I\pm)}:\quad & \eta= \rho \frac{|w|}{a} \sin (\pm a \ln (\rho) + b), \quad a> 0,\ b \in \R,\\
\label{II} {\rm (II\pm)}:\quad & \eta= \rho |w| ( \pm \ln (\rho) + b),\quad b \in \R,\\
\label{III} {\rm (III\pm)}:\quad & \eta= \rho \frac{|w|}{a} \sinh (\pm a \ln (\rho) + b), \quad a> 0,\ b \in \R,
\end{align}
and the twist potential is always,
\be
\omega = wz,\quad w\neq 0.
\ee
The solutions extending to infinity are (\ref{II}) and (\ref{III}) with the $+$ sign (note that they are positive after some $\rho$). For the case (\ref{III},+), the metric components $V$, $W$, $\eta$ and $e^{2\gamma}/\eta$ get the form,
\begin{gather}
V =  \frac{2a}{|w|} e^{-b} \rho^{1-a},\quad W =   \text{s}(w)e^{-b} \rho^{1-a}, \label{V_of_rho_Lewis}\\
\eta = \frac{|w|}{2a} (e^{b}\rho^{1+a}-e^{-b}\rho^{1-a}),\quad \frac{e^{2\gamma}}{\eta} = c\rho^{(a^2-1)/2},
\label{VsolT2} 
\end{gather}
where $\text{s}(w)=w/|w|$ is the sign of $w$ and $b\in \mathbb{R}, c>0$ and $a>0$ are free parameters. The angular velocity function $\Omega$ is,
\begin{equation}
\Omega =  \frac{2a}{w}\frac{e^{-b}\rho^{-a}}{(e^{b}\rho^{a}-e^{-b}\rho^{-a})},
\end{equation}
and note that $\Omega$ is set to be zero at infinity. Not all of the solutions can model the asymptotic of coaxial arrays of black holes like the ones we are considering. Concretely, if $a\geq 1$ the lapse function for the hypersurface $t=0$ tends to zero as infinity something that can be ruled out by the maximum principle on the lapse equation taking into account that the lapse is zero on the horizons. As $a$ tends to zero, $\eta$ in (III+) degenerates into (II+). The asymptotic models are therefore (III+) with $0<a<1$ and (II+).   

The metrics of the models (III+) are Kasner to leading order (except for the cross term $-2\text{s}(w)e^{-b}\rho^{1-a}dtd\phi$). Recalling that the Kasner metrics have the form,
\begin{equation} \label{kasnersol}
g_{K} = -e^{-b}\rho^{\alpha} dt^2 + e^{b}\rho^{2-\alpha}d\phi^2 + c\rho^{\alpha^{2}/2-\alpha} \left( d\rho^2 + dz^2 \right),
\end{equation}
we see that in case (III+) we have $\alpha = 1-a$. Thus, the Kasner exponent $\alpha$ of the models (III+) is restricted to vary in $(0,1)$. 

\subsubsection{Absence of angle defects in our periodic set up} \label{absence_of_struts}

Co-axial multi-black hole configuration can display angle defects on the axis sometimes called struts, (see for instance \cite{Wei90} and references therein). MKN solutions have no struts due to the periodicity and this is also the case for the configurations we are considering here. Let's show this, so assume $(\sigma,\omega)$ is a solution with the symmetries we have considered, $\sigma$ even and $\omega$ odd on $[0,\infty)\times [-L/2,L/2]$.  

Let $q = \gamma - \sigma - \ln \rho$. Then,
\begin{equation} \label{quad_q}
q_\rho = (\sigma_\rho^2 + \sigma_z^2) + \frac{\rho}{4 \eta^2} (\omega_\rho^2 - \omega_z^2) ,\quad q_z = \frac{\rho}{2} (\sigma_z \sigma_\rho + \frac{1}{\eta^2} \omega_z \omega_\rho) .
\end{equation}
Define, 
\begin{equation} \label{definition_Delta_q}
\Delta q := q \mid_{\Ax_+} - q \mid_{\Ax_-},
\end{equation}
which is the difference in the value of $q$ on the two components of the axis. Absence of struts implies $\Delta q = 0$. As $q$ is constant over the horizon, it suffices to show $q(L/2)=q(-L/2)$. The integral of the closed form,
\begin{equation}
\left( \frac{\rho}{4} (\sigma_\rho^2 + \sigma_z^2) + \frac{\rho}{4 \eta^2}
(\omega_\rho^2 - \omega_z^2)\right) d\rho + \frac{\rho}{2} (\sigma_z \sigma_\rho + \frac{1}{\eta^2} \omega_z,
\omega_\rho) dz 
\end{equation}
on the segment from $(1,L/2)$ to $(1,L/2)$ is zero by the symmetries of $\sigma$ and $\omega$. Also for the symmetries, the integral on the segments $[0,1]\times \{L/2\}$ and $[0,1]\times \{-L/2\}$, oriented in the same direction are equal. Therefore the integral on the three consecutive intervals is zero, so $\Delta q = q(L/2)-q(-L/2)=0$.

\subsection{Boundary conditions on the rectangle}\label{sec_boundary_conditions}

From now it will be more convenient to work with $\sigma=\ln \eta - 2\ln\rho$ (this choice has some advantages, see \cite{DO, Frolov2003, Wei90}). In terms of $(\sigma,\omega)$ the harmonic map equations \eqref{REE10} and \eqref{REE20} become,
\begin{align}
\Delta_{\rho} \sigma &= - \frac{e^{-2 \sigma} | \partial \omega |^2}{\rho^4},
\label{REE1_sigma} \\
\Delta_{\rho} \omega &= 4 \frac{\langle \partial \omega,  \partial \rho\rangle}{\rho} + 2\langle
\partial \omega, \partial \sigma\rangle. 
\label{REE2_omega}
\end{align}
The boundary conditions for this system will be the same as the boundary condition for the harmonic map heat flow, and are as follows.
\vspace{.2cm}

\noindent {\it Boundary conditions on $\mathcal{A}$ and $\mathcal{H}$}. Recall that $\mathcal{H}=\{-m<z<m\}$ and that the axis $\mathcal{A}$ has two components in $[-L/2,L/2]$, $\mathcal{A}_{+}=\{m\leq z\leq L/2\}$ and $\mathcal{A}_{-}=\{-L/2\leq z\leq -m\}$. 

The boundary conditions for $\omega$ are,
\begin{align}
& \omega\big|_{\mathcal{A}_{+}} = 4J,\quad \omega\big|_{\mathcal{A}_{-}}=-4J,\\ 
& \partial_{\rho} \omega\big|_{\mathcal{H}} = 0.
\end{align}
The first, Dirichlet condition, comes from fixing the angular momentum of the horizons to $J$. The second, Neumann condition, can be obtained by demanding the spacetime smoothness of $\Omega=W/\eta$ that is linked to $\omega$ by \eqref{quadOmega}.  

The boundary conditions for $\sigma$ are,
\begin{align}
& \partial_{\rho} \sigma \big|_{\mathcal{A} \setminus \partial \Ho}=0,\\ 
& \partial_{\rho} (\sigma + 2\ln \rho)\big|_{\mathcal{H}} = 0,\\ 
& \lim_{(\rho,z) \rightarrow (0, \pm m)} \frac{\sigma}{\sigma_0} = 1,
\end{align} 
where $\sigma_0$ is the reference Kerr solution given $A,J$ and $m(A,J)$, where  
\begin{equation}
m = \sqrt{\frac{A}{16 \pi}} \frac{  1 - \left( 8 \pi J/ A \right)^2 }{\sqrt{ 1 + \left( 8 \pi J/ A \right)^2 } }.
\end{equation}
This enforces the area of the solution to be actually $A$, noting that the area and temperature are given by
\begin{align*}
A &= 2\pi \int_{-m}^m e^{\gamma} dz, \\
\kappa &= e^{-\gamma}
\end{align*}
\vspace{.2cm}



\noindent {\it Boundary conditions at $\rhomax$}. Since we expect Lewis asymptotic and the Komar angular momentum is $J$, for the boundary condition of $\omega$ we simply set, 
\begin{equation}
\omega(\rhomax,z)= (8J/L) z. 
\end{equation}

The boundary condition for $\sigma$ is delicate and we motivate it as follows. Given a solution to (\ref{REE10})-(\ref{REE20}) one can get a Smarr type of expression for the Komar mass (per black hole) for the Killing field $\partial_{t}$ by integrating the Komar form on any 2-torus $t=0$ and $\rho={\rm const}$ of the spacetime, of course after identifying $z=-L/2$ to $z=L/2$. The expression is,
\begin{equation}\label{Smarr}
M(\rho) = \frac{1}{4} \int_{-L/2}^{L/2} (- \rho \partial_\rho \sigma + \Omega \partial_z \omega ) dz,
\end{equation}
which is in fact $\rho-$independent.

For the Lewis solutions (III+) we have $\Omega\rightarrow 0$ as $\rho\rightarrow \infty$, $\rho\partial_{\rho}\sigma\rightarrow a-1$ as $\rho\rightarrow \infty$, $\partial_{z}\omega=w$, so we obtain,
\begin{equation} \label{mass_T2_sym}
M = \frac{L}{4} (1-a).
\end{equation}
Thus, the Kasner exponent $\alpha=1-a$ is equal to $4M/L$.

The Lewis solution (III+) on the finite rectangle $[0,\rhomax]\times [-L/2,L/2]$ satisfies the decay 
\begin{equation} \label{Neumann_like_asymptotic}
\partial_\rho \sigma \big|_{\rhomax} = \frac{4 M}{L \rhomax} + o(1/\rhomax),\quad \rhomax \gg 1
\end{equation} 
where $M$ is the value of the Smarr mass, still unknown.

Now, since we are evolving a harmonic map heat flow, at any time $\tau \geq 0$ the integral in \eqref{Smarr} associated to $(\sigma,\omega)$ at $\tau$ is not necessarily constant as a function of $\rho$, and even the function $\Omega$ is not well defined since the integrability conditions (\ref{quadOmega}) do not necessarily hold. Therefore, we have to give a definition for $M(\rho,\tau)$ so that we can use the approximation \eqref{Neumann_like_asymptotic} as boundary condition. By integrating by parts the last term in \eqref{Smarr}, and using \eqref{quadOmega}, we have a candidate

\begin{equation}
M(\rho, \tau) = 2 \Omega(\rho, \tau) J \bigg|_{z=-L/2} - \frac{1}{4} \int_{\rho} ( \rho \partial_\rho \sigma + \frac{\rho}{\eta^2} \partial_\rho \omega \omega ) dz,
\end{equation}
where $\Omega(\rho, \tau)$ is prescribed to be

\begin{equation}
\Omega(\rho, \tau) = \int_{\rhomax}^\rho \frac{\rho'}{\eta^2} \partial_z \omega d \rho',
\end{equation}
with the functions evaluated at $z= - L/2$. If $(\sigma,\omega)$ converges to a solution, then $M(\rho , \tau)$ converges to a constant function. Taking this into account, a natural dynamical condition arise if we take the $\rho$-\textit{average} of $M(\rho,\tau)$ over the numerical domain, at each time step, and
impose 
\begin{equation} \label{mean_M}
M(\rhomax) = \overline{M}(\tau),
\end{equation}
which in particular links the values of $M$ near the axis with those values far
away.\footnote{In general we have a weighted average, which range from being
completely uniform on the interval $(0 , \rhomax)$ to associating more
weight to the region near the axis and cutting off the region past certain
$\rho_0 < \rhomax$. Different weights were explore in the numerical
simulations.} Then, our boundary condition for $\sigma$ at $\rhomax$ is
given by
\begin{equation} \label{asymptotic_condition_sigma}
\partial_\rho \sigma \big|_{\rhomax} = - \frac{4 \overline{M}(\tau)}{L \rhomax}.
\end{equation} 
Observe that in this condition there is no explicit intervention of the total
angular momentum, and there is also no reference to any specific asymptotic
model. This is an important characteristic of this boundary condition: it will
be the same for any periodic configuration, as long as the
solutions become $z-$independent asymptotically.

It remains to fix the boundary conditions for the integration of the quadratures \eqref{quadOmega} and \eqref{quad_q}. For $\Omega$, we impose 

\begin{equation}\label{bc_Omega}
\Omega \big|_{\rhomax} = 0,
\end{equation}
which is the natural condition for the asymptotic models, and also consistent with the definition of $M$. The boundary condition on $q$ is just that it vanishes at just one point in $\mathcal{A}_{0}$. As it was presented above, the symmetries on $\sigma$ and $\omega$ and the integrability equations (\ref{quad_q}) imply this same condition holds at any other point of any axis component $\mathcal{A}_{i}$, preventing struts in between black holes.

\subsection{Initial data for the harmonic map heat flow}
 
We need some initial condition at $\tau=0$ of the heat flow, which we will call \textit{seed}. In particular, it is desirable that the seed contains the prescribed singular behaviour of the solutions at the horizons, so that we can define a well-posed numerical problem without singularities. We decompose $\sigma$ and $\omega$ as follows: we split them as a sum of known solutions to the non-periodic problem plus a perturbation $\bar{\sigma}, \bar{\omega}$. In the case a single horizon per period, this sum is constructed in the same fashion as the function $\sigma_{MKN}$ was defined: following \cite{DO}, let $\sigma_0 (\rho,z)$ and $\omega_0 (\rho ,z)$ be the solutions to the asymptotically flat Kerr black hole with momentum $J$ and horizon $\Ho_0$, and define
\begin{equation}\label{sig_om_desc}\begin{split}
\sigma =& \sigma_0 + \sigma_r + \bar{\sigma} \\
\omega =& \omega_0 + \omega_r + \bar{\omega},
\end{split}
\end{equation}
where,
\begin{align}
\sigma_r (\rho , z) &= C +  \sum_{n=1}^{\infty} \left( \sigma_0(\rho,z - nL ,J)  + \sigma_0(\rho,z + nL,J) - \frac{4M}{nL} \right) \nonumber \\
\omega_r (\rho , z) &= \sum_{n=1}^{\infty} \left( \omega_0(\rho,z - nL ,J)  + \omega_0(\rho,z + nL - D/2 ,J) \right). \nonumber,
\end{align}
where $C$ is a constant such that $\sigma_r \mid_{\partial \Ho} = 0$, that is,
its value at the poles is zero. Here the constants $\frac{4M}{nL}$ are needed
for the series to be convergent (since asymptotically, each term goes as $-
\frac{2M}{\sqrt{(x-nL)^2 + \rho^2}}$ and therefore we need to cancel out this
divergent term, as in \cite{KN}). In our actual numerical calculations we use a
cut-off value $N_d$ for $n$, which we take $N_d \gg 1$, which can be thought of
as the number of ``domains'' we stack on both the top and below the central
domain.

We expect $(\bar{\sigma},\bar{\omega})$ to be regular throughout the evolution.
By inserting the decomposition \eqref{sig_om_desc} into \eqref{REE1_sigma} and
\eqref{REE2_omega}, and using the fact that $(\sigma_0 ,\omega_0)$ is a solution, we
obtain the evolution equations for $\bar{\sigma}$ and $\bar{\omega}$,
\begin{equation}\label{bar_sigma_eq}\begin{split}
\dot{\bar{\sigma}} &= \Delta \bar{\sigma} + \Delta \sigma_r + \frac{e^{-2 \sigma_0} |\partial \omega_0 |}{\rho^4} \left( e^{-2 (\bar{\sigma} + \sigma_r)  } -1 \right) \\
&\quad + \frac{e^{-2 (\sigma_0 + \sigma_r + \bar{\sigma}) }}{\rho^4}
\left( |\partial \omega_r|^2 + |\partial \bar{\omega}|^2 + 2 \left( \partial_i
\omega_r \partial^i \omega_0 + \partial_i \omega_r \partial^i \bar{\omega} +  \partial_i \bar{\omega} \partial^i \omega_0 \right) \right),
\end{split}
\end{equation}
\begin{equation}\label{bar_omega_eq}\begin{split}
\dot{\bar{\omega}} &=  \Delta \bar{\omega} + \Delta \omega_r - \frac{4}{\rho} \left( \partial_\rho \omega_r + \partial_\rho \bar{\omega} \right) - 2 \left( \partial_i \omega_0 \partial^i \sigma_r + \partial_i \omega_0 \partial^i \bar{\sigma}   \right. \\
&\quad \left. + \partial_i \omega_r \partial^i \sigma_0 + \partial_i \omega_r \partial^i \sigma_r + \partial_i \omega_r \partial^i \bar{\sigma} + \partial_i \bar{\omega}  \partial^i \sigma_0 + \partial_i \bar{\omega} \partial^i \sigma_r + \partial_i \bar{\omega} \partial^i \bar{\sigma} \right).
\end{split}
\end{equation}
These are the equations that we solve numerically, with the following
boundary conditions, which can be read off from the conditions for the total
functions $(\omega,\sigma)$,
\begin{equation}\label{bomega_BC}
\bar{\omega} (\rhomax) = 0, \quad \partial_\rho\bar{\omega} \mid_{\Ho} = 0, \quad \bar{\omega} \mid_{\Ax} = 0,
\end{equation}
since $\omega_0 + \omega_r$ already satisfies the asymptotic linear behaviour
for $\omega$, and 
\begin{equation}\label{bsigma_BC_axis}
\partial_\rho\bar{\omega} \mid_{(\Ax \cup \Ho) \setminus \partial \Ho} = 0, \quad \bar{\sigma} \mid_{\partial \Ho} = 0,
\end{equation}
with the asymptotic condition 

\begin{equation} \label{bsigma_asymptotic_condition} 
\partial_\rho \bar{\sigma} \big|_{\rhomax} = - \frac{4 \overline{M(\rho)}^\rho}{L \rhomax} - \overline{\partial_\rho (\sigma_0 + \sigma_r)}^z \big|_{\rhomax}
\end{equation} 
The last term is not $z-$independent, since the series are truncated, and
therefore we take its average on $z$, denoting by $\overline{X}^x$ the average
along $x$ coordinate of the variable $X$. We will call $\beta$ the dinamical
quantity given by the right hand side of equations
\eqref{bsigma_asymptotic_condition}.

\section{Numerical Implementation}\label{sec_numerical_implementation}

We use a grid adapted to a finite computational region where the
Weyl-Papapetrou coordinates range as follows,
\[
(\rho,z) \in [0, \rhomax]\times [-L/2, L/2].
\]
We use $N_\rho+1$-point Chebyshev grid to discretize $\rho$ and a uniform grid of
$N_z$ points, which are semi displaced with respect to the boundaries $z=\pm L/2,$
to discretize $z$. Along this section we use sub indices to identify grid points
and grid values:
\begin{equation}\label{grid}\begin{split}
\rho_i &= \frac{1}{2}\rhomax \left( 1 - \cos \Bigl( \frac{\pi}{N_\rho}i
\Bigr) \right) \quad i = 0,...,N_\rho,\\ 
z_j &= -\frac{L}{2} + \frac{L}{N_z} \Bigl( j + \frac{1}{2} \Bigr) , \quad j =
0,...,N_z-1.
\end{split}
\end{equation}
Observe that the symmetry axis, $\{ \rho=0\}$, is included in the grid while the
axis $\{ z = 0\}$ is not. Also, the $z$-grid is defined in such a way that the
poles $\mathcal{H}\cap \mathcal{A}$ are at the middle of two consecutive grid
points. Derivatives with respect to $\rho$ are approximated by the derivatives
of the polynomial interpolation on the Chebyshev grid, while derivatives with
respect to $z$ are approximated as the derivatives of the standard Fourier
interpolation on the uniform grid. This is, we use pseudo spectral and spectral
collocation methods in $\rho$ and $z$ respectively. 

We wrote two independent versions of Python codes to carry out the
numerical computations to cross check the results. The implementation of the spectral method
is through the standard {\tt rfft} routines provided by {\tt NumPy}, while for
the pseudo spectral derivatives and integrals we tried various matrix
implementations \cite{dmsuite, trefethen, elbarbary} that produce no significant
differences between them. The values of the analytical Kerr solution and their
derivatives were obtained as Python codification using {\tt SymPy} and {\tt
Maple}.

As explained in section \ref{sec_boundary_conditions}, every solutions to our
problem is obtained as the final state of the parabolic flow
\eqref{bar_sigma_eq},\eqref{bar_omega_eq}.  The singularities of $\sigma$ at
$\Ho$, and the starting point for the evolution of the parabolic flow, are
handled via the splitting of $\sigma$ and $\omega$ as in equation
\eqref{sig_om_desc} with the introduction of the seed $\sigma_0 + \sigma_r$ and
$\omega_0 + \omega_r$. This is, the initial value for $\bsigma$ and $\bomega$ is
always taken as zero, and the boundary conditions are given by equations
\eqref{bomega_BC}--\eqref{bsigma_asymptotic_condition}.

A particular numerical problem is defined once the values of physical
parameters: $J$, $A$, $L,$ the value of $N_d$ which amounts the number of
periods we stack to build the seed, and the values of the numerical parameters:
$\rhomax$, $N_\rho$ and $N_z,$ are chosen. As explained before, we choose the
values of $L$ judiciously so that the poles fall at the middle of two
consecutive gridponts at $\rho=0$. In section \ref{sec_results} we show the
precise values used in our runs.

The time evolution for the parabolic flow is implemented with Euler's method. One could argue that
Euler's method is a low precision method, but it is explicit, simple to
implement, and more important we are only seeking
for the final stationary solutions of the equations where all time variations go
to zero together with the associated truncation error. We are not interested in
the precision along the time evolution. Near the stationary state the truncation
error is dominated by that of the space discretization. Of course the time step
is subordinated to the grid sizes so as to obtain a numerically stable scheme at
all times. The closer to the symmetry axis, the stricter the CFL condition
becomes, since the Chebyshev mesh size is smaller and the derivatives of various functions
involved are larger. In most of our runs a time step $\delta t= 10^{-4}$
turned out to be suitable.

\paragraph{The boundary conditions.} Let us denote, as usual, the grid functions
as
\[
\bsigma_{i,j}(t) = \bsigma(\rho_i, z_j, t), \quad \mbox{and} \quad
\bomega_{i,j}(t) = \bomega(\rho_i, z_j, t), \quad i=0, \dots, N_\rho,\quad j =
0, \dots, N_z-1,
\]
and the pseudo spectral derivative matrix associated to the Chebyshev
$\rho$-grid as
\[
D_{i,k}, \quad i,k = 0, \dots, N_z-1.
\]

Given the grid functions at time $t$, a single Euler step
determines both grid functions at time $t+\delta t$ in all gridpoints with $1
\le i\le N_\rho -1.$ Periodicity in $z$ is an intrinsic part of the
implementation. The values at $i=0$ (axis and horizon) and $i=N_\rho$ (outer
boundary) at time $t+\delta t$ are determined by the boundary conditions as
follows.
\begin{enumerate}
\item $\bomega_{N_\rho,j}(t+\delta t) = 0$ for all $j$ (homogenous Dirichlet
condition for $\bomega$ at $\rhomax$).
\item $\bomega_{0,j}(t+\delta t) = 0$  for all $j$ such that $z_j \in {\cal A}$ (homogenous Dirichlet
condition for $\bomega$ at the axis). 
\item Solve $\sum_{k=1}^{N_\rho} D_{0,k} \bomega_{k,j}(t+\delta t) = 0$ for
$\bomega_{0,j}(t+\delta_t)$, for those $j$  such that $z_j \in {\cal H}$
(homogeneous Neumann condition for $\bomega$ on the horizon). 
\item $\bsigma_{0,j}(t+\delta t)$ and $\bsigma_{N_\rho,j}$ are determined by solving a
$2\times 2$ system that implements the homogeneous Neumann condition for $\bsigma$ at the
axis and the dynamical inhomogeneous Neumann condition for $\bsigma$ at the outer
boundary $\rho=\rhomax.$ For each value of $j$ the system is
\[
\left(\begin{array}{cc}
D_{0,0} & D_{0,N_\rho}\\
D_{N_\rho,0} & D_{N_\rho,N_\rho}
\end{array}\right)\begin{pmatrix}
\bsigma_{0,j}(t+\delta t) \\ \bsigma_{N_\rho,j}(t+\delta t)
\end{pmatrix} = \begin{pmatrix}
c \\ d
\end{pmatrix}
\]
where the inhomogeneity is 
\[\begin{split}
c &= -\sum_{i=1}^{N_\rho-1} D_{0,i}\bsigma_{i,j}(t+\delta t),\\
d &= -\sum_{i=1}^{N_\rho-1} D_{N_\rho,i}\bsigma_{i,j}(t+\delta t) + \beta,
\end{split}
\]
where $\beta$ is the dynamical value given by the right hand side of equation
\eqref{bsigma_asymptotic_condition}.
\item Finally, to keep the homogeneous Dirichlet condition for $\bsigma$ at
the poles we compute the (very small) violation $\bsigma_{\rm pole}$ as the
average of the $\bsigma(t+\delta t)$ values at the two nearest neighbour grid points
on $\rho=0$ to any of the poles, and subtract this value from $\bsigma(t+\delta t)$ on the
whole grid.
\end{enumerate}

The evolution of the parabolic flow approaches the stationary state only in an
asymptotic manner. To meassure the distance to stationarity, we compute the $L_2$
norm of the right hand side equations \eqref{bar_sigma_eq},\eqref{bar_omega_eq}.
This is an absolute meassure of the ``error''. Then, we compute the relative
errors
\begin{equation}\label{convergence_error}
\varepsilon_{\bar \sigma}=\frac{\|\mbox{rhs}(\bar\sigma)\|}{\|\bar\sigma\|}, \quad
\varepsilon_{\bar\omega} = \frac{\|\mbox{rhs}(\bar\omega)\|}{\|\bar\omega\|},
\quad \varepsilon_{\sigma}=\frac{\|\mbox{rhs}(\bar\sigma)\|}{\|\sigma\|},
\quad \mbox{and} \quad \varepsilon_{\omega} = \frac{\|\mbox{rhs}(\bar\omega)\|}{\|\omega\|}.
\end{equation}

Finally, the quadratures for the various functions that need to be done are
implemented by standard spectral {\tt rfft} routines along $z$ direction and
Clenshaw-Curtis integration along $\rho$ direction.

\section{Results}\label{sec_results}

In this setion we present two series of simulations computed with our code for
two different values of angular momentum: $J=1/4$ and $J=1/2$. For the series
with $J=1/4$ we present and analyse in detail several aspects of the solutions
obtained. For the series with $J=1/2,$ not to be redundant, we simply show a
table with some relevant quantities computed. We choose to
compute solutions whose horizon's area is $A=16\pi.$ Thus, recalling that $\kappa$ is the Kerr temperature given $A$ and $J$, the horizon's
semi-length becomes
\[
m = \frac{4-J^2}{2\sqrt{4+J^2}}.
\]
In the cases we study, $m = 0.9095$ for $J=1/2$ and $m=0.9768$ for $J= 1/4$. The various solutions in each series correspond to different values of the
parameter $L,$ that we choose as
\[
L = \frac{N_z}{N_h}m,
\]
where $N_h$ is the number of $z$-gridpoints inside each horizon. In all cases
the computational domain has $\rhomax = 40,$ and the computing grid is defined
with $N_\rho = 79$ and $N_z = 100$ (see \eqref{grid}). 

\subsection{First series: $J=1/4$} 

\begin{figure}[t]
\captionsetup{margin=1cm}
\begin{center}
\includegraphics[width=7cm]{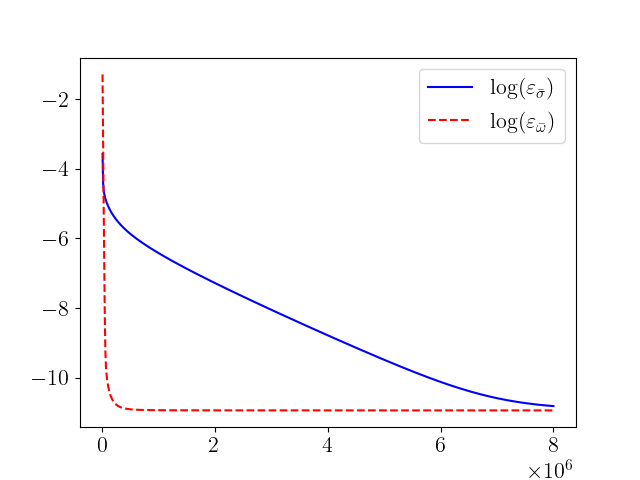}
\includegraphics[width=7cm]{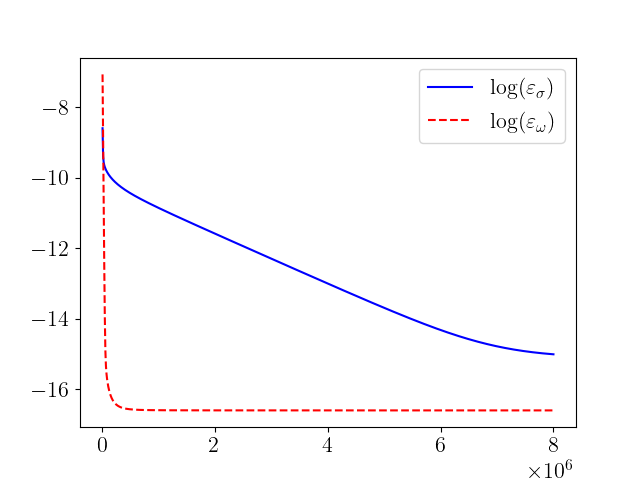}
\caption{Convergence of the parabolic flow to stationary state as a function of time steps for
the solution with $N_h=40$ in Table \ref{table_conv_series_1}.}\label{fig_convergence_stat}
\end{center}
\end{figure}

\begin{figure}[h!]
\captionsetup{margin=1cm}
\begin{center}
\includegraphics[width=7cm]{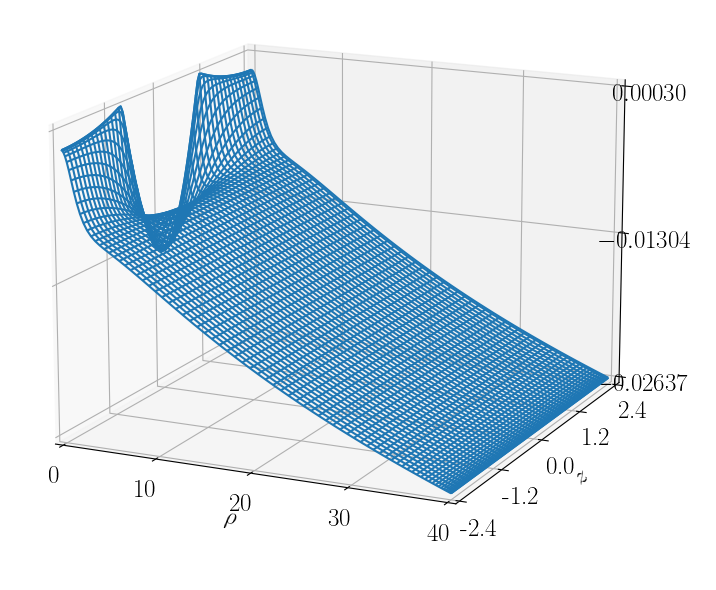}
\includegraphics[width=7cm]{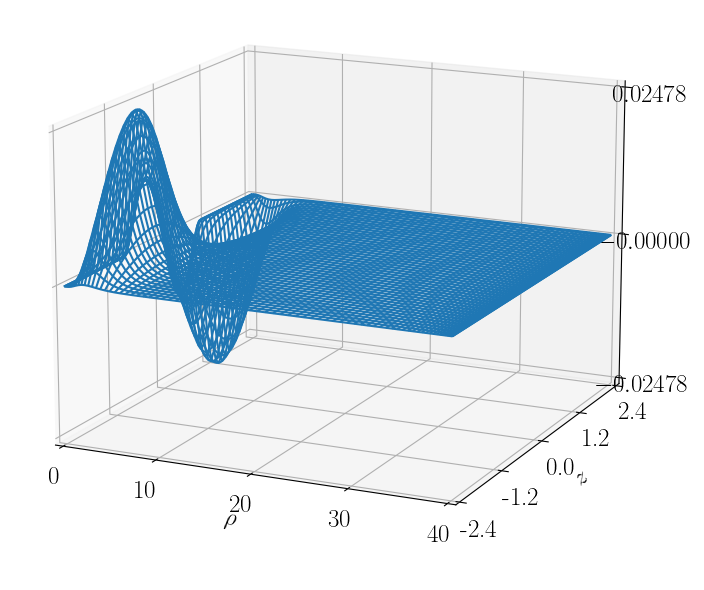}
\includegraphics[width=7cm]{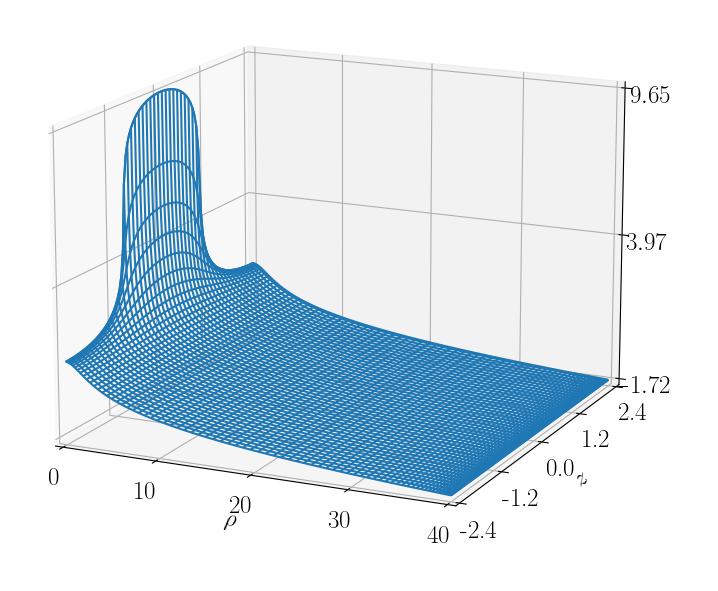}
\includegraphics[width=7cm]{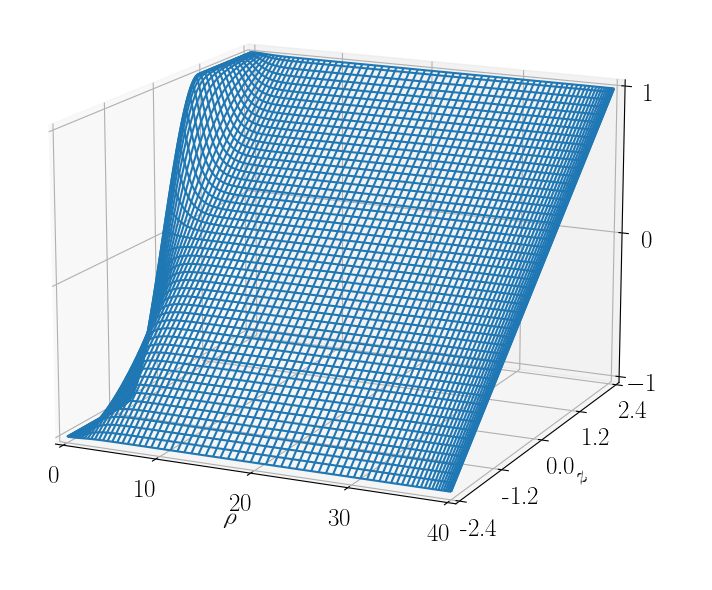}
\end{center}
\caption{Plots of the solution corresponding to $L=4.8839$ of Table
\ref{table_conv_series_1}. From left to right, from top to bottom: $\bsigma$,
$\bomega$, $\sigma$, $\omega$.}\label{fig_sol_Nh_40}
\end{figure}

\paragraph{Convergence of the parabolic flow and regularity of the solution.} The convergence of the parabolic
flow to stationary state turns out to be slow. For all the solutions in this
series, we stopped the flow after computing $8\times 10^6$
steps with $\delta t = 10^{-4}$, where we found that the relative errors
$\varepsilon_{\bsigma}$ and $\varepsilon_{\bomega}$ are comparably small.
Typical plots of $\varepsilon_\bsigma$ and $\varepsilon_\bomega$, in logarithmic scale, along the
evolution of the flow is shown in Figure \ref{fig_convergence_stat}. 
Also, the regularity of the solution at this final time is checked by computing
$\Delta q,$ as defined in \eqref{definition_Delta_q}, by path integrating around
the horizon from just above the upper pole to just below the lower pole. As it
can be seen, the violation of regularity turns out to be extremely small. The
final values of the relative errors for all the runs in this series, together
with the values of $\Delta q$ are shown in table \ref{table_conv_series_1}.
\begin{table}[h!]
\captionsetup{margin=1cm}
\begin{center}
{\footnotesize
\begin{tabular}{ccccccc}
$N_h$ & $L$ & $\varepsilon_\bsigma$ & $\varepsilon_\bomega$ & $\varepsilon_\sigma$ & $\varepsilon_\omega$ & $\Delta q$ \\
\hline 
22 & 8.8798 & 2.19$\times 10^{-6}$ & 1.40$\times 10^{-6}$ & 7.94$\times 10^{-8}$ & 3.22$\times 10^{-9}$ & -6.66$\times 10^{-15}$ \\
28 & 6.9770 & 2.21$\times 10^{-5}$ & 4.83$\times 10^{-6}$ & 1.31$\times 10^{-7}$ & 1.29$\times 10^{-8}$ & -4.17$\times 10^{-14}$ \\
34 & 5.7457 & 2.46$\times 10^{-5}$ & 1.06$\times 10^{-5}$ & 2.34$\times 10^{-7}$ & 3.25$\times 10^{-8}$ & -5.10$\times 10^{-14}$ \\
40 & 4.8839 & 2.01$\times 10^{-5}$ & 1.78$\times 10^{-5}$ & 3.04$\times 10^{-7}$ & 6.19$\times 10^{-8}$ & -1.50$\times 10^{-13}$ \\
46 & 4.2468 & 1.59$\times 10^{-5}$ & 2.62$\times 10^{-5}$ & 3.84$\times 10^{-7}$ & 1.03$\times 10^{-7}$ &  2.49$\times 10^{-14}$ \\
50 & 3.9071 & 1.47$\times 10^{-5}$ & 3.27$\times 10^{-5}$ & 4.89$\times 10^{-7}$ & 1.40$\times 10^{-7}$ &  4.39$\times 10^{-14}$ \\
52 & 3.7568 & 1.47$\times 10^{-5}$ & 3.63$\times 10^{-5}$ & 5.77$\times 10^{-7}$ & 1.62$\times 10^{-7}$ &  4.59$\times 10^{-14}$ \\
54 & 3.6177 & 1.52$\times 10^{-5}$ & 4.01$\times 10^{-5}$ & 7.07$\times 10^{-7}$ & 1.87$\times 10^{-7}$ & -2.07$\times 10^{-13}$ \\
56 & 3.4885 & 1.63$\times 10^{-5}$ & 4.42$\times 10^{-5}$ & 9.04$\times 10^{-7}$ & 2.16$\times 10^{-7}$ & -8.26$\times 10^{-14}$ \\
58 & 3.3682 & 1.83$\times 10^{-5}$ & 4.87$\times 10^{-5}$ & 1.22$\times 10^{-6}$ & 2.48$\times 10^{-7}$ & -1.32$\times 10^{-13}$ \\
60 & 3.2559 & 2.17$\times 10^{-5}$ & 5.36$\times 10^{-5}$ & 1.75$\times 10^{-6}$ & 2.85$\times 10^{-7}$ & -2.35$\times 10^{-13}$ \\
\hline
\end{tabular}
\caption{Relative error after $8\times 10^6$ time steps and violation of
regularity, $\Delta q,$ for the solutions in the series.}\label{table_conv_series_1}
}
\end{center}
\end{table}
We choose the run with $L=4.8839$ (corresponding to $N_h=40$) as an example to show the plots of
the solution and relevant functions, Figure \ref{fig_sol_Nh_40} shows the plots
of $\bsigma$, $\bomega$, $\sigma$ and $\omega$.

\paragraph{Mass, angular velocity and Kasner parameter.} At the final time we
compute, for every solution, several relevant quantities. The mass $M$ is
computed as the integral $M(\rhomax)$. The horizon's angular velocity is obtained
as the averaged value of $\Omega(\rho=0)$ in one of the horizons.\footnote{The
computation of $\Omega$ is singular at the horizon; we compute $\Omega$ strictly
in the interior and get the value of $\Omega(\rho=0)$ by simple linear
extrapolation from the first and second internal gridpoints.}. We also compute
the Kasner exponent in two different ways: the first value is obtained from the
Mass, as $4M/L$,
while the second value is obtained from the asymptotic behaviour of the function
$V$; more precisely, as the slope of a linear regression of $\ln(V)$ as function
of $\ln(\rho)$ in the {\em asymptotic region} of the computational domain, see \eqref{V_of_rho_Lewis}. We
arbitrarily define the {\em asymptotic region} of the domain as the portion of
the domain, adjacent to $\rhomax,$ corresponding to 30\% of $\rho$ grid points.
The two values obtained for the Kasner exponent are in very good agreement.
All these quantities, for the solutions in this series are shown in Table
\ref{table_results_series_1}.
\begin{table}[h!]
\captionsetup{margin=1cm}
\begin{center}
{\footnotesize
\begin{tabular}{cccccc}
$L$ & $M$ (mass) & Angular velocity &  $\alpha$ (from $M$) & $\alpha$ (from $V$) \\
\hline
8.8798 & 1.0095 & 6.5753$\times 10^{-2}$ & 4.5476$\times 10^{-1}$ & 4.5477$\times 10^{-1}$ \\
6.9770 & 1.0119 & 7.0513$\times 10^{-2}$ & 5.8014$\times 10^{-1}$ & 5.8017$\times 10^{-1}$ \\
5.7457 & 1.0164 & 7.9507$\times 10^{-2}$ & 7.0758$\times 10^{-1}$ & 7.0768$\times 10^{-1}$ \\
4.8839 & 1.0250 & 9.6689$\times 10^{-2}$ & 8.3947$\times 10^{-1}$ & 8.3977$\times 10^{-1}$ \\
4.2468 & 1.0422 & 1.3130$\times 10^{-1}$ & 9.8167$\times 10^{-1}$ & 9.8264$\times 10^{-1}$ \\
3.9071 & 1.0640 & 1.7495$\times 10^{-1}$ & 1.0893 & 1.0916 \\
3.7568 & 1.0807 & 2.0831$\times 10^{-1}$ & 1.1506 & 1.1541 \\
3.6177 & 1.1036 & 2.5413$\times 10^{-1}$ & 1.2202 & 1.2257 \\
3.4885 & 1.1358 & 3.1881$\times 10^{-1}$ & 1.3024 & 1.3113 \\
3.3682 & 1.1831 & 4.1348$\times 10^{-1}$ & 1.4050 & 1.4201 \\
3.2559 & 1.2557 & 5.5907$\times 10^{-1}$ & 1.5426 & 1.5698 \\
\hline
\end{tabular}
\caption{Relevant quantities computed for the solutions in the series.}\label{table_results_series_1} 
}
\end{center}
\end{table}

\begin{figure}[h!]
\captionsetup{margin=1cm}
\begin{center}
\includegraphics[width=7cm]{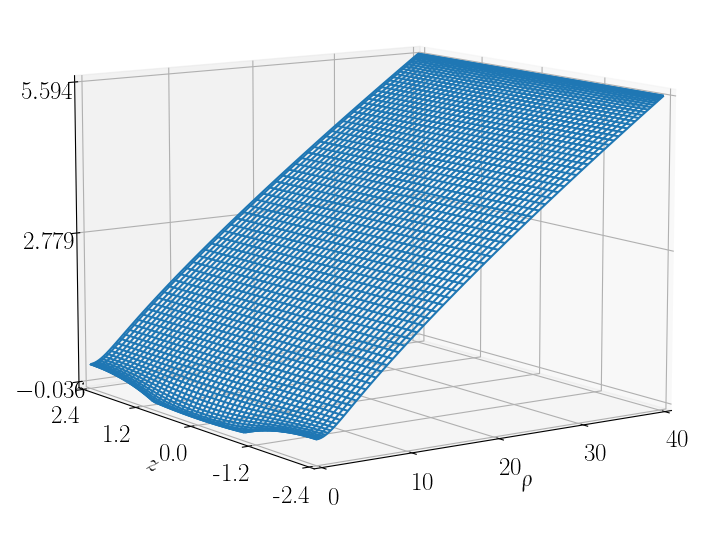}
\includegraphics[width=7cm]{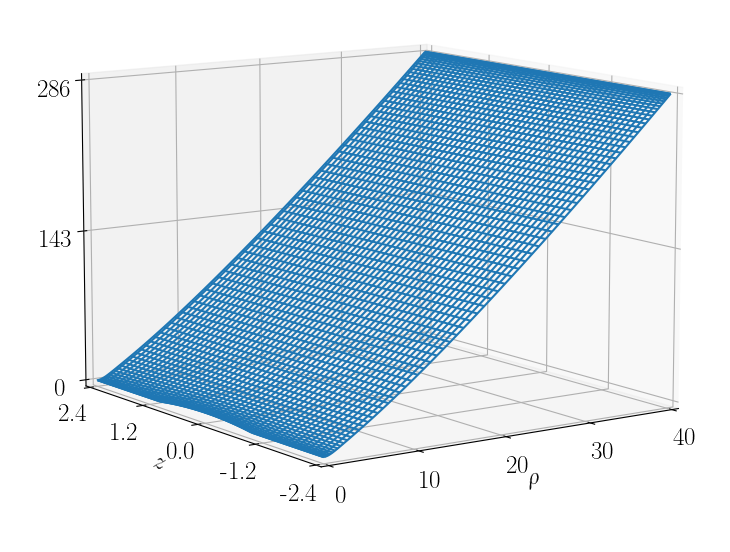}
\includegraphics[width=7cm]{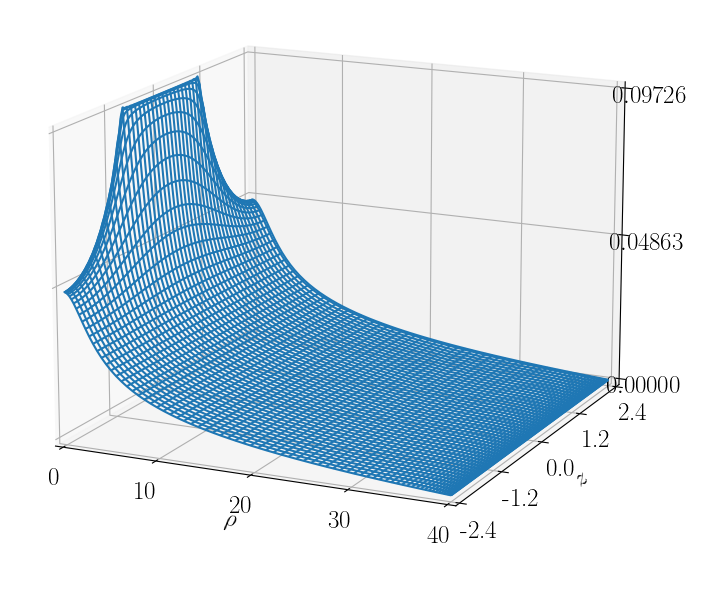}
\includegraphics[width=6cm]{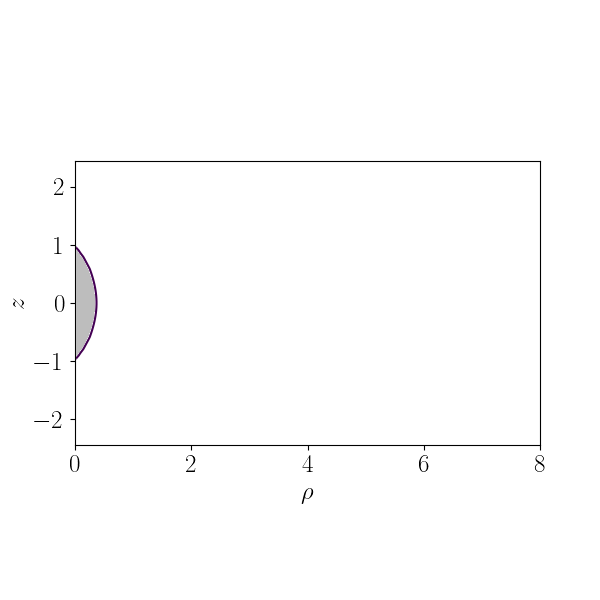}
\end{center}
\caption{Plots of the metric functions $V$ (top left) and $\eta$ (top right),
and plot of $\Omega$ and the ergosphere (as a gray region) at the bottom, for the solution with $N_h=40$ of Table
\ref{table_conv_series_1}.}\label{fig_metric_Nh_40}
\end{figure}

\begin{figure}[h!]
\captionsetup{margin=1cm}
\begin{center}
\includegraphics[width=4.5cm]{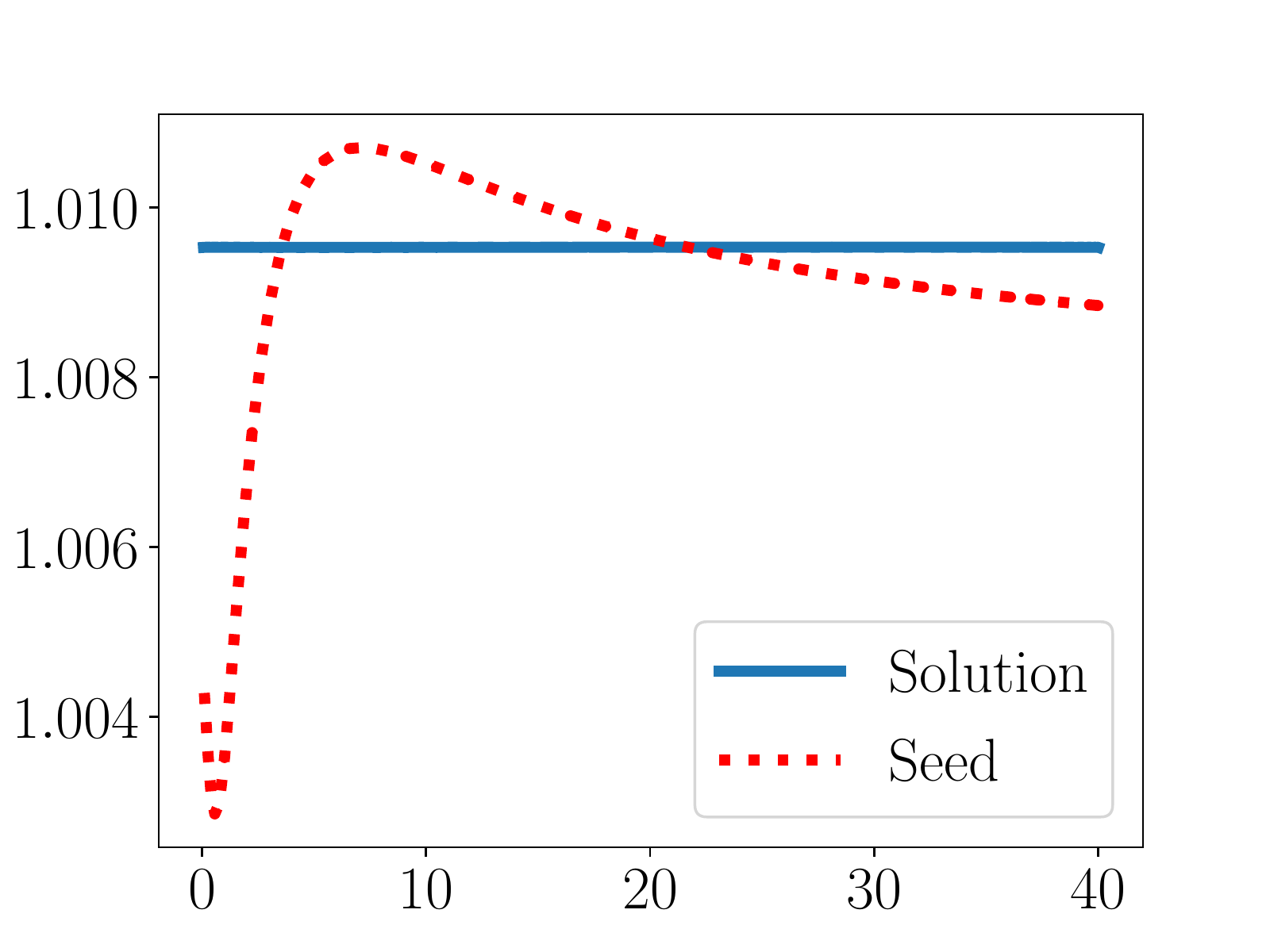}
\includegraphics[width=4.5cm]{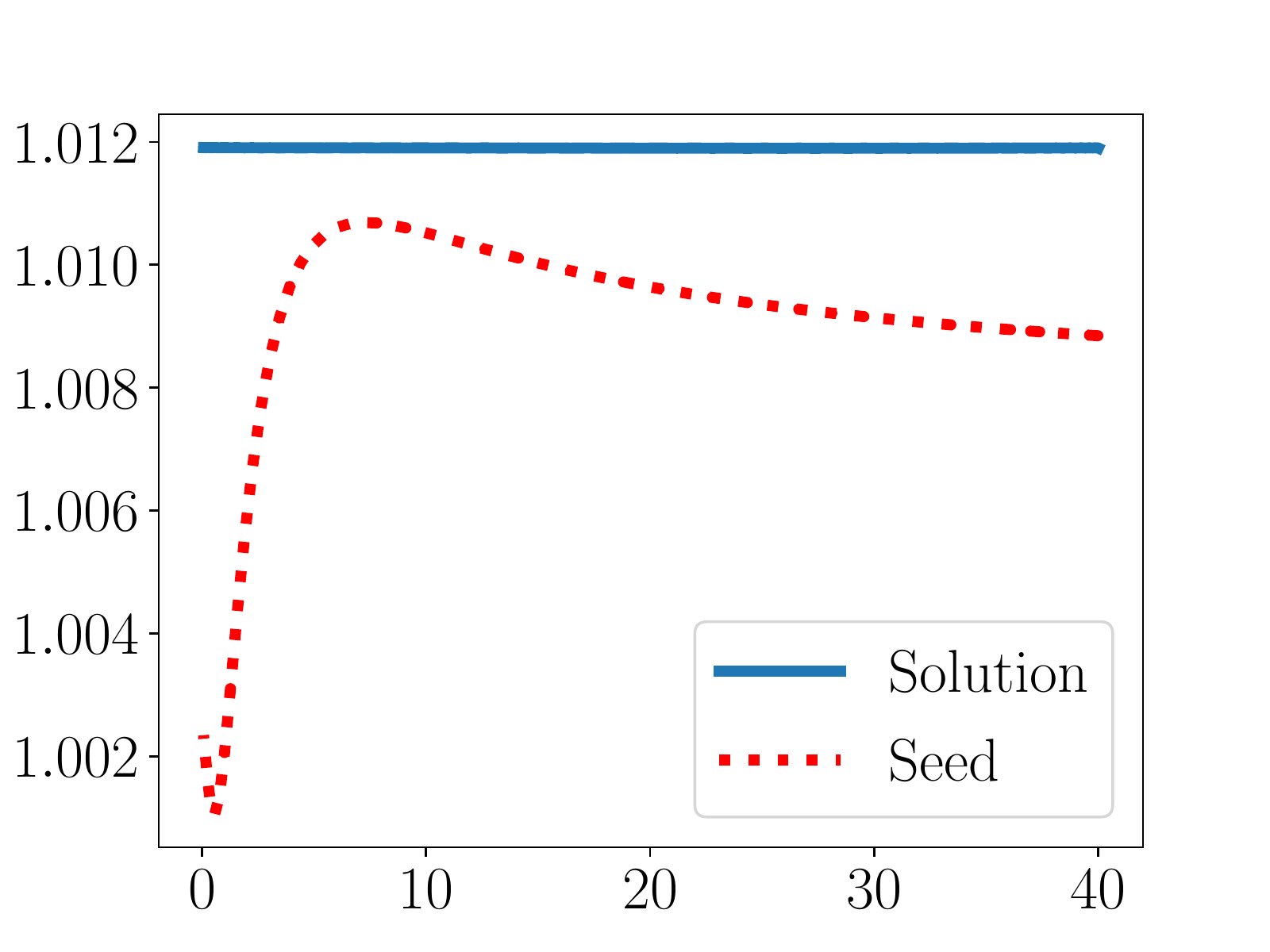}
\includegraphics[width=4.5cm]{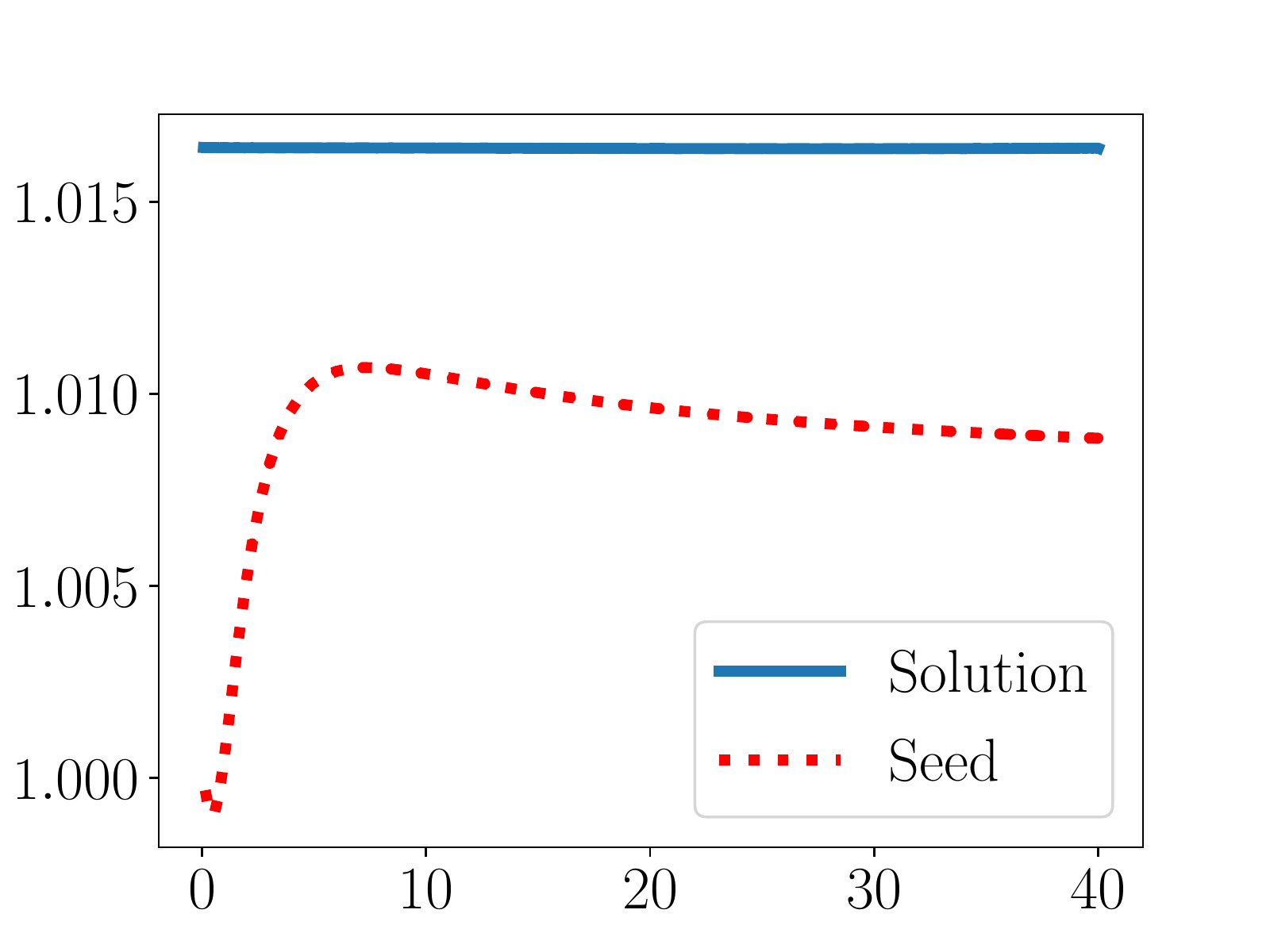}\\
\includegraphics[width=4.5cm]{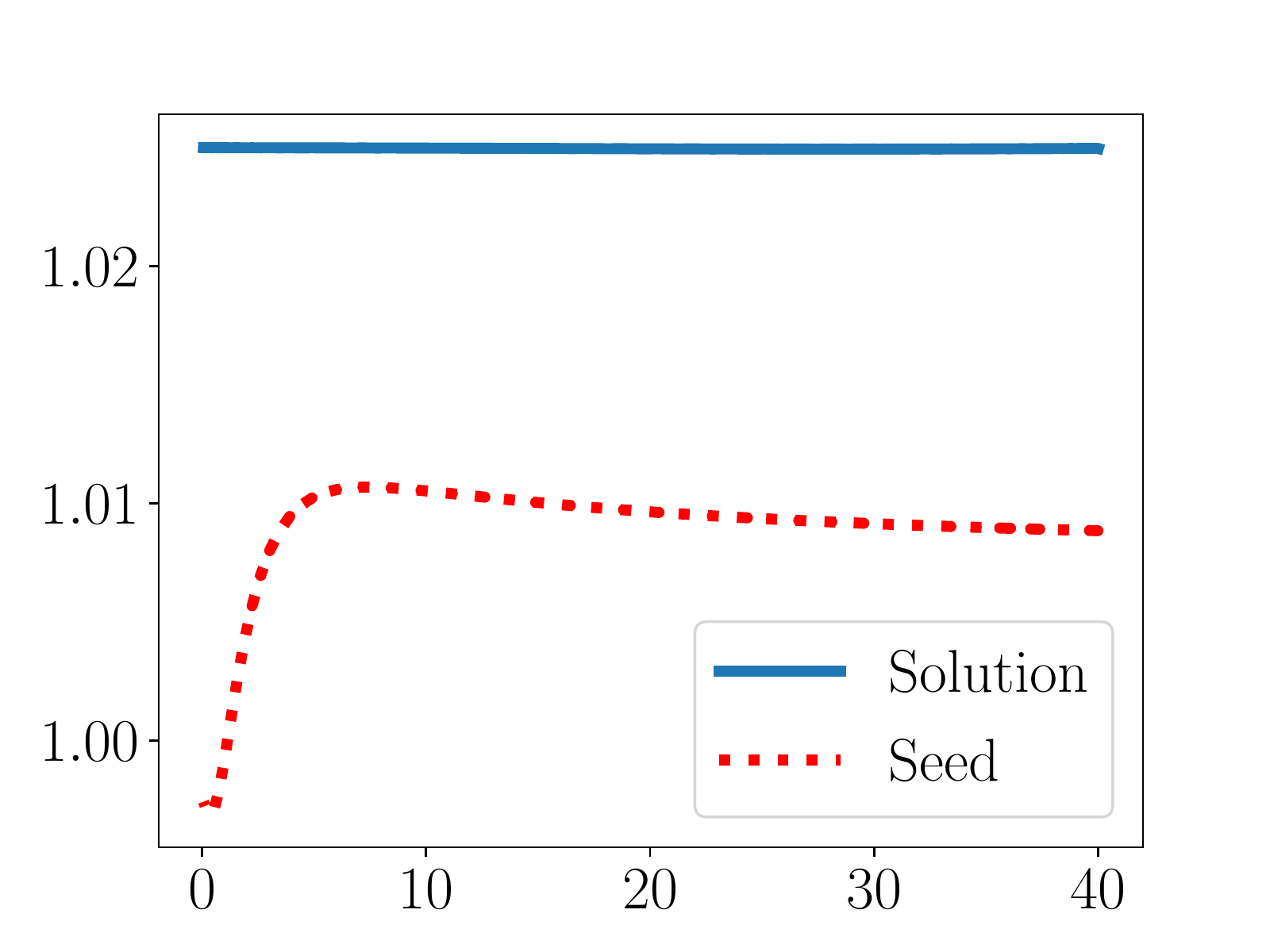}
\includegraphics[width=4.5cm]{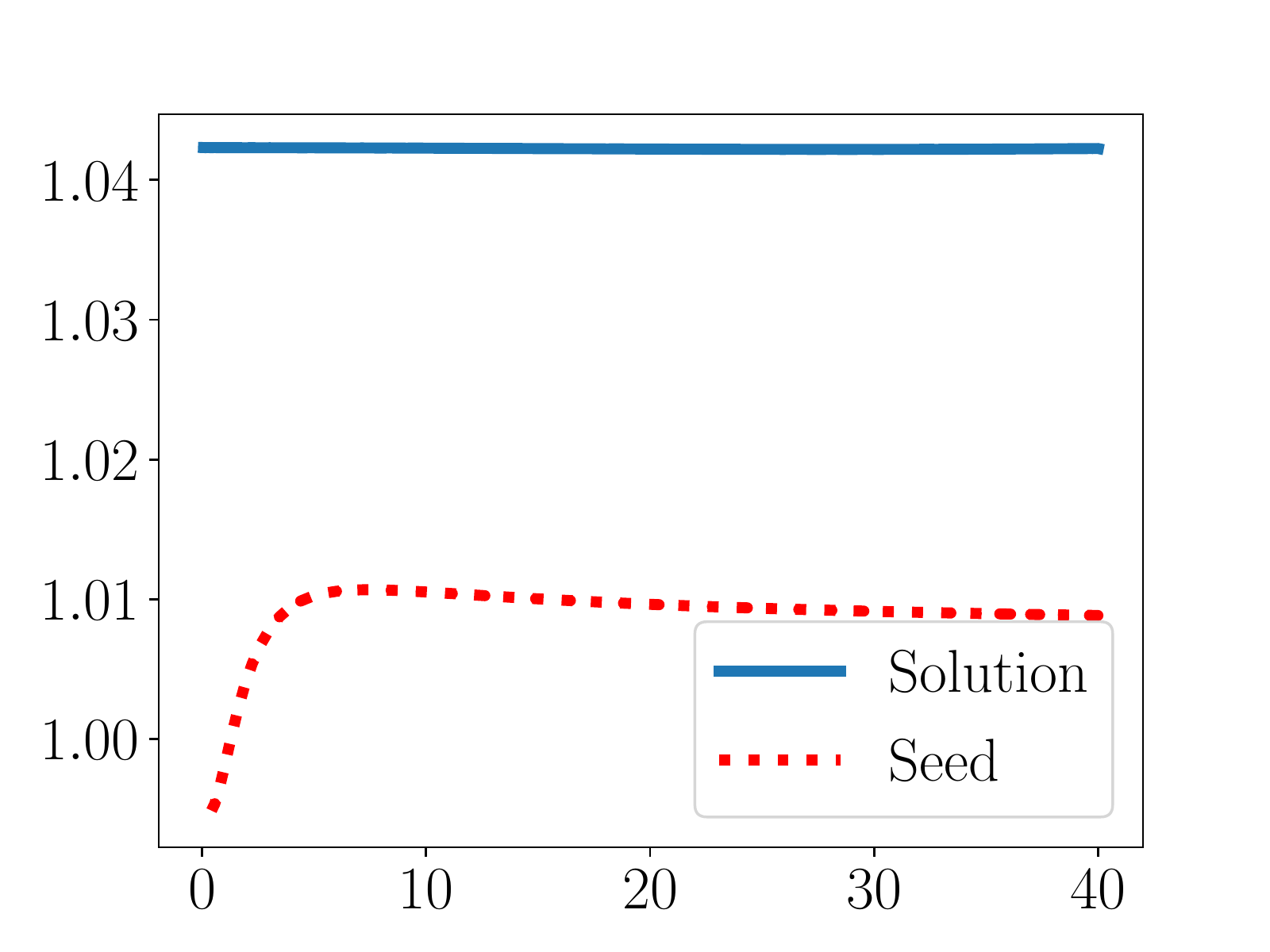}
\includegraphics[width=4.5cm]{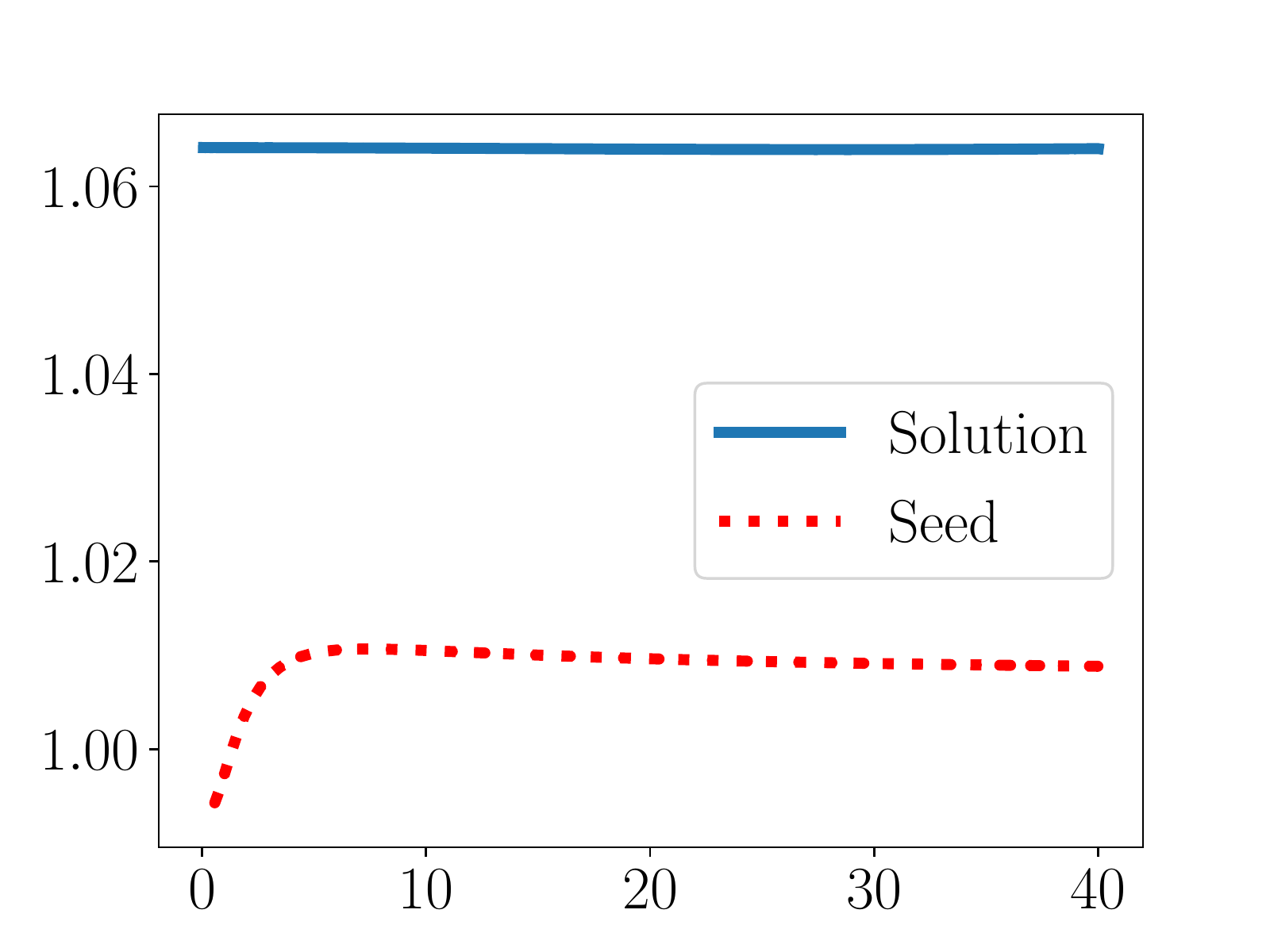}
\end{center}
\caption{Plots of $M(\rho)$ for the seed and for the numerically computed
solution compared for six of the solutions in the series. From left to right
from top to bottom the plots correspond to the cases with $Nh =22, 28, 34, 40, 46, 50$ in Table
\ref{table_conv_series_1}.}
\label{fig_smarr_series_1}
\end{figure}

Figure \ref{fig_metric_Nh_40} shows the plots of relevant metric
functions obtained for the solution corresponding to $N_h = 40.$ 
\paragraph{The Smarr identity.} We found that an very sensitive test for
checking the computations is the validity of the Smarr identity, this is, the
constancy of $M(\rho)$ (see equation \ref{Smarr}) as a function of $\rho$.  In
this sense the plot of $M(\rho)$ became crucial to test to correctness of the
outer boundary condition for $\bsigma.$ In Figure \ref{fig_smarr_series_1} we
show plots that compare $M(\rho)$ computed for the seed (initial data for the
flow) and $M(\rho)$ at final time for the six numerically computed solutions with larger
$L$ in the series.

\begin{figure}[h!]
\captionsetup{margin=1cm}
\begin{center}
\includegraphics[width=7cm]{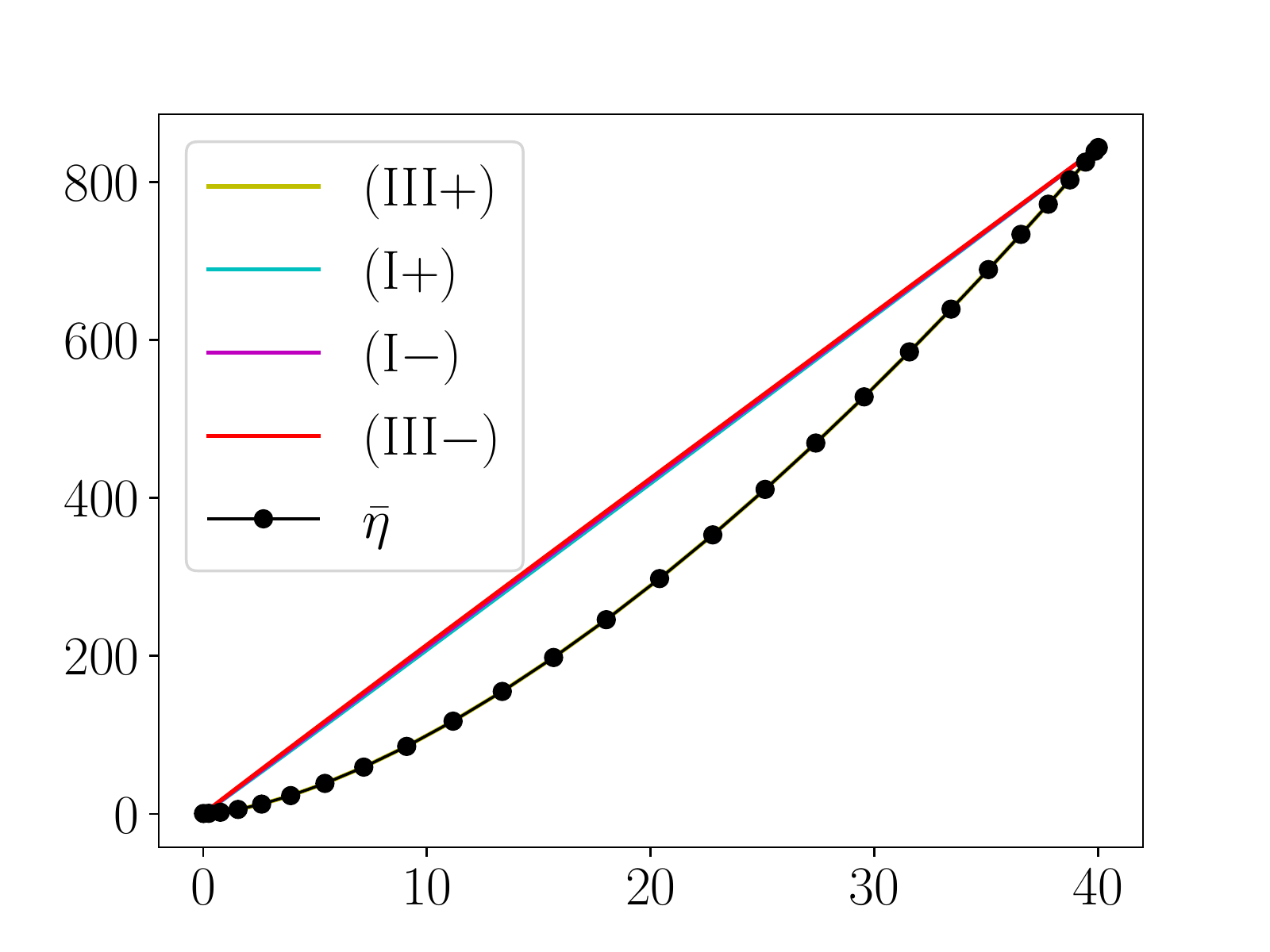}
\includegraphics[width=7cm]{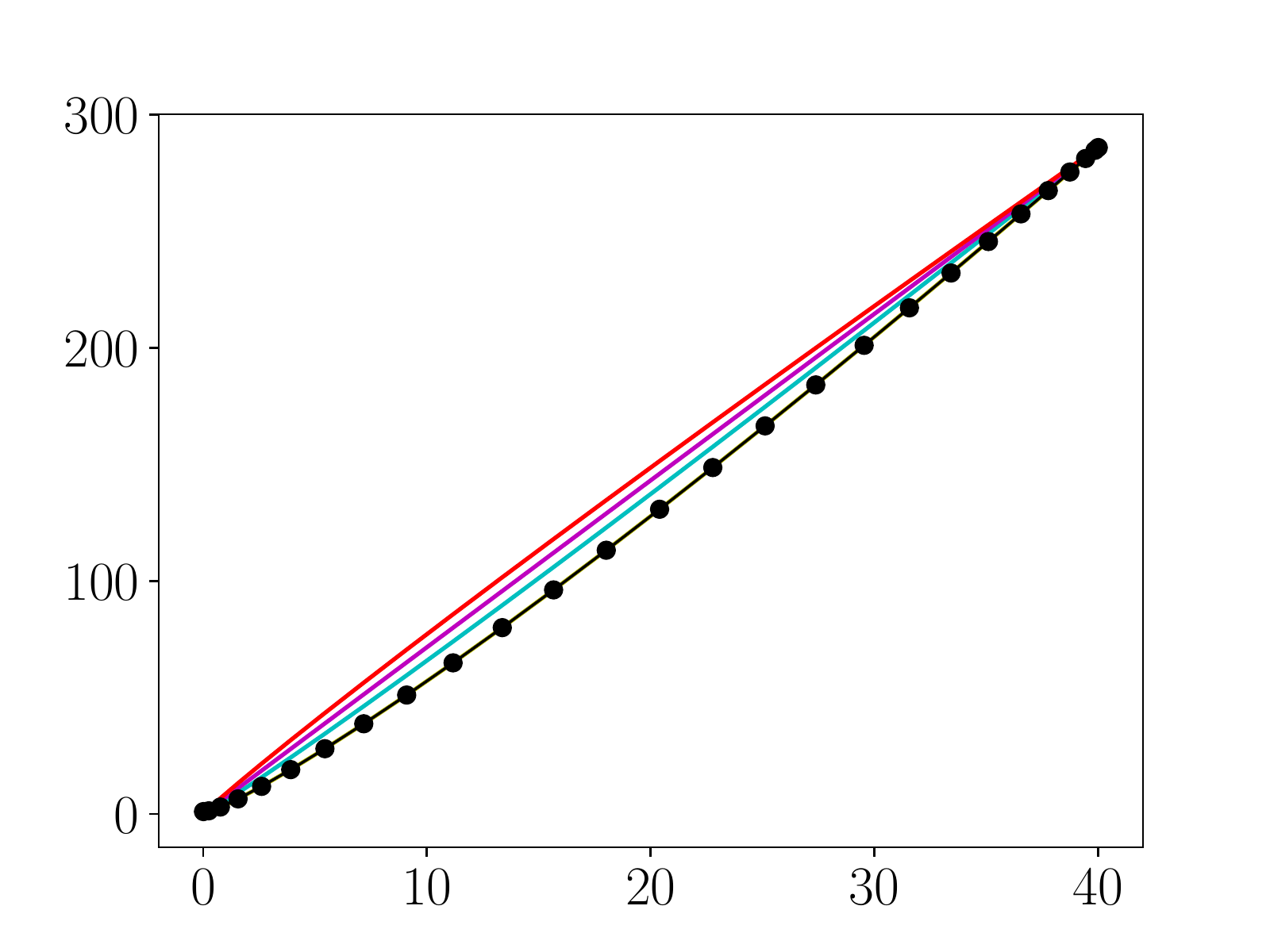}\\
\includegraphics[width=7cm]{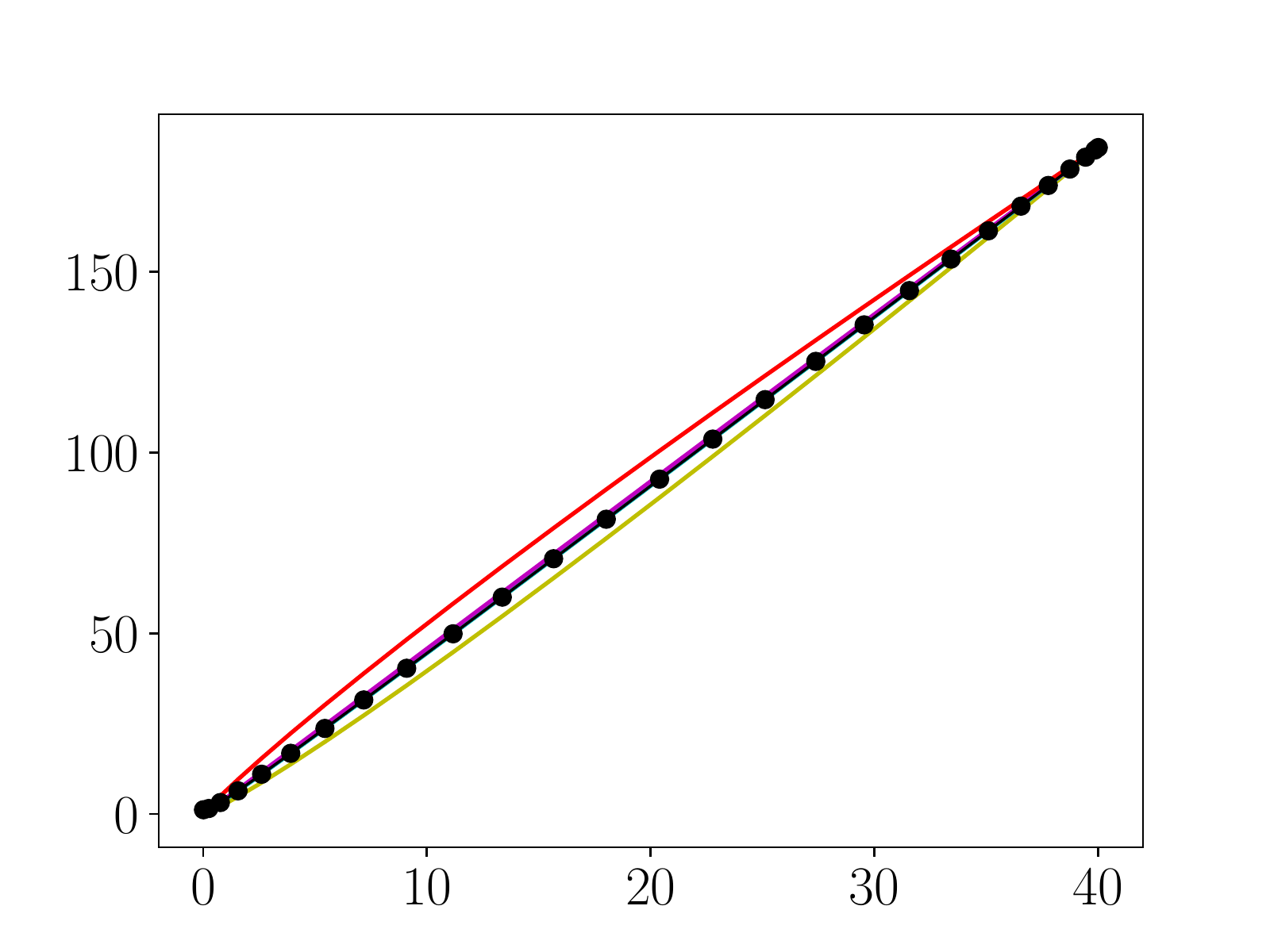}
\includegraphics[width=7cm]{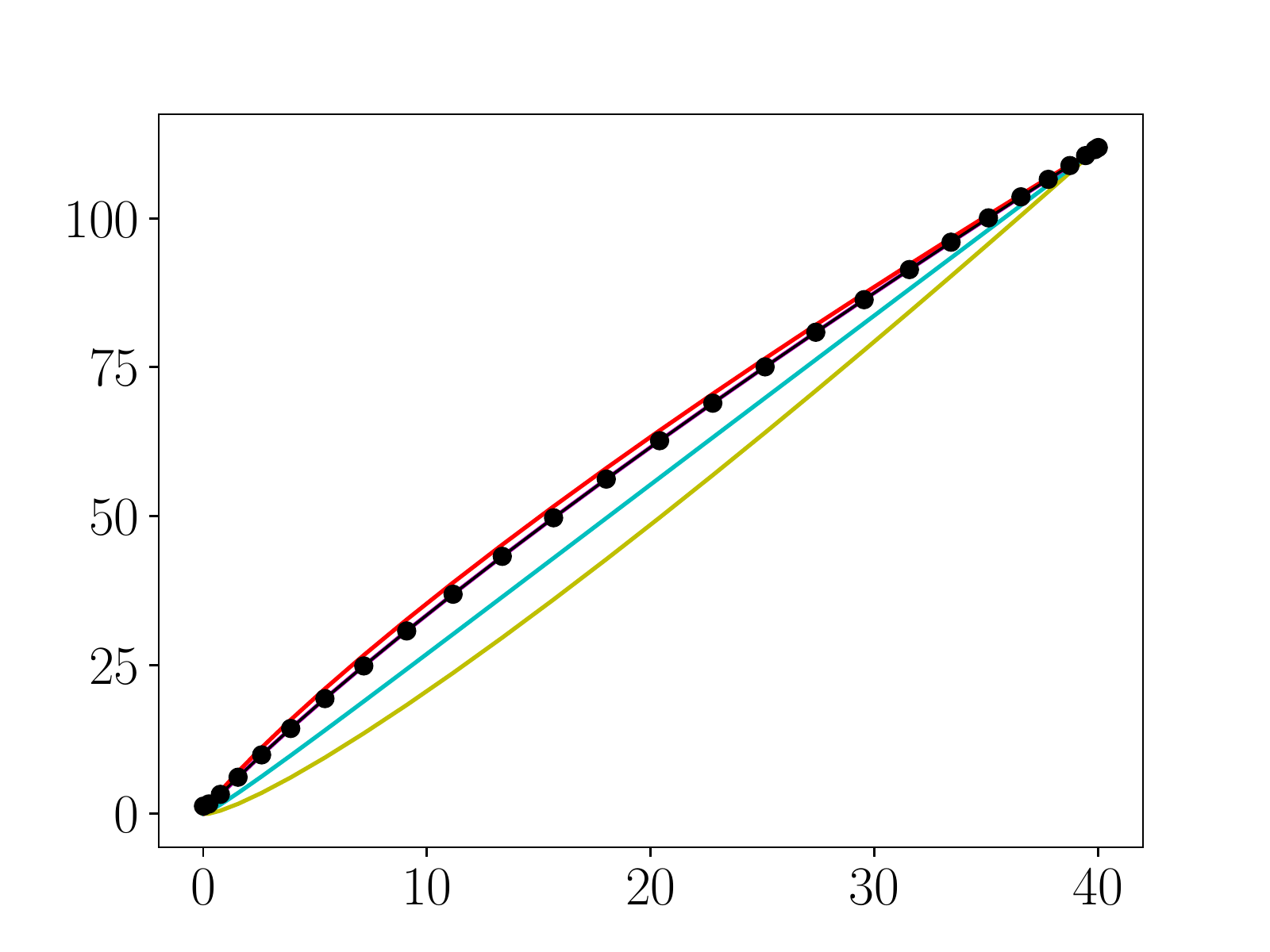}\\
\includegraphics[width=7cm]{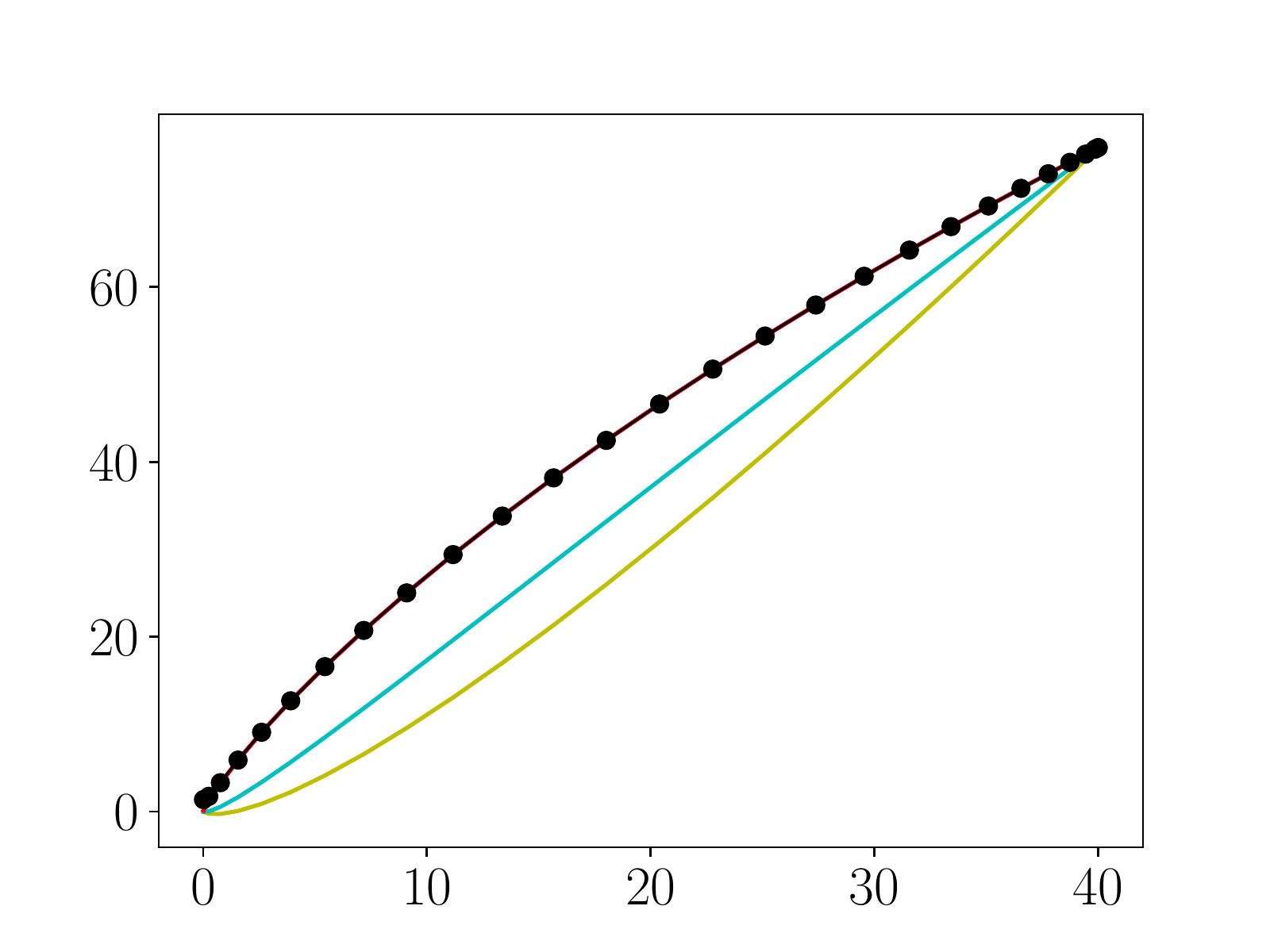}
\includegraphics[width=7cm]{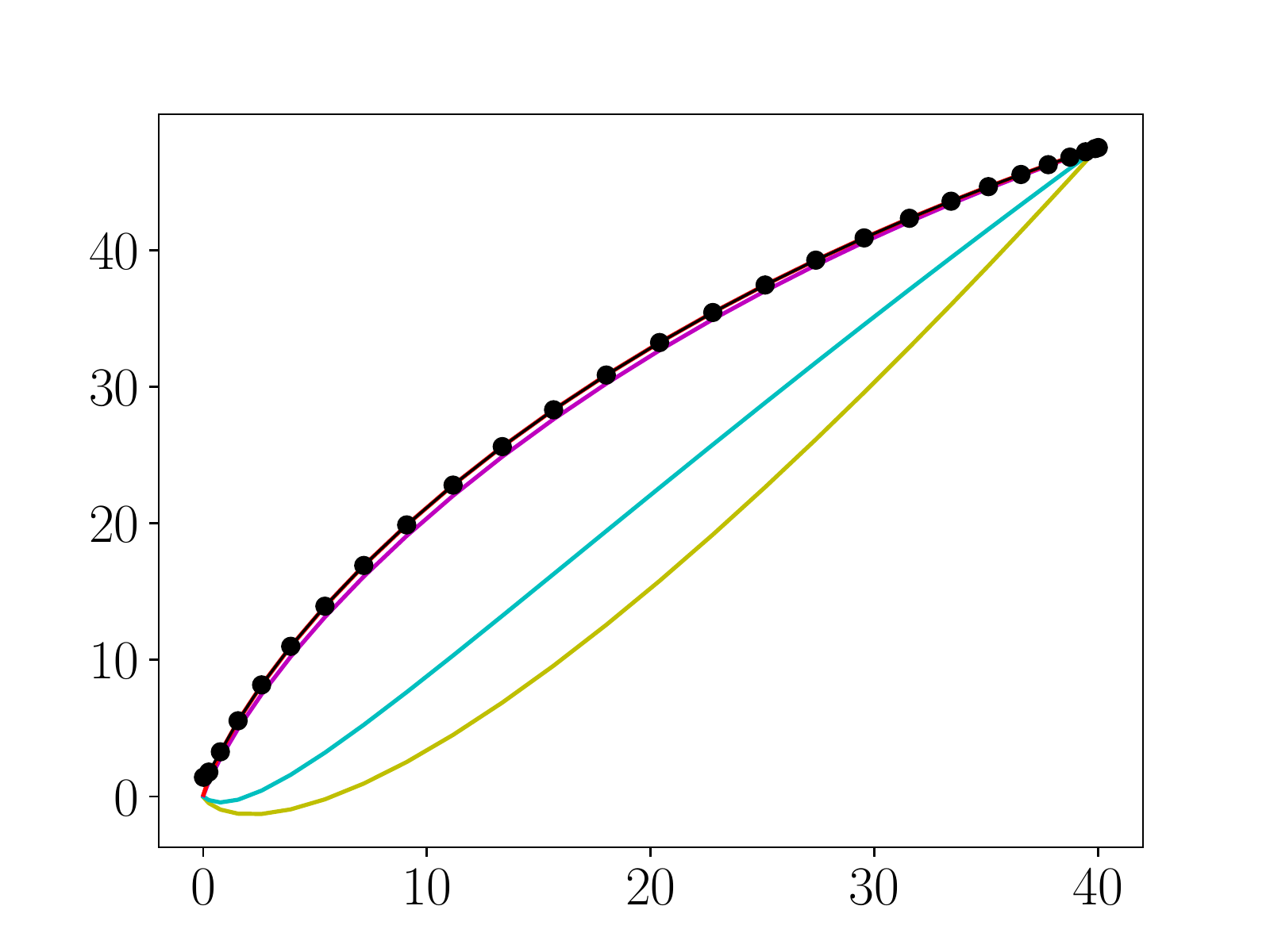}
\end{center}
\caption{Plots of $\bar\eta$ together with the fitting models (III$+$), (I$+$),
(I$+$) and (III$+$) for the six solutions in Table \ref{table_fitting_etas}
corresponding, from left to right, from top to bottom, to $L=8.8798,~ 4.8839,~
4.2468,~ 3.7568,~ 3.4885,~ 3.2559$. The $\bar\eta$ curve overlaps the best
fitting model in all cases, while some of the bad fitting model curves also
overlap, that is why no all curves are visible in all
plots.}\label{fig_fitting_etas}
\end{figure}

\paragraph{Best fitting asymptotic models} We want to check which of the
asymptotic candidate solutions fits
better the numerically computed solution. To this end, we take the average on
the $z$ axis for the function $\eta$ as to get a $z$ independet function $\bar
\eta(\rho).$ We then compute the best fitting model $\eta$-function for
the six posibilities given by models (I$+$), (I$-$), (II$+$), (II$-$), (III$+$)
and (III$-$) (see equations \eqref{I}, \eqref{II}, and
\eqref{III}). To do this we minimize the
deviation of the model $\eta(a,b,\rho)$ from $\bar\eta$ by  varying the
parameter $a$, or taking $a=0$ for the models (II), and choosing
$b$ in such a way that $\eta$ functions are coincident at the outer boundary. We meassure 
the mentioned deviation by computing the integrated square difference in the asymptotic region:
\[
\Delta \eta = \int_{\mbox{asympt. region}} \Bigl(\bar\eta(\rho) -
\eta(a,b,\rho)\Bigr)^2~ d\rho.
\]
The results of fitting the eleven solutions in Table \ref{table_conv_series_1} are shown in Table \ref{table_fitting_etas}.
\begin{table}[h!]
\captionsetup{margin=1cm}
\begin{center}
{\footnotesize
\begin{tabular}{c|ccc|cc|ccc}
& (III$+$) &&& (II$+$) && (I$+$) &&\\ 
$L$ & $a$ & $b$ & $\Delta\eta$ & $b$ & $\Delta\eta$ & $a$ & $b$ & $\Delta\eta$ \\
\hline
8.8798 &	0.5451 &	2.6149	 & 2.4$\times 10^{-7}$ &	89.9297 & 5.2$\times 10^1$ &	0.0001 &    0.0090 &	5.2$\times 10^1$  \\
6.9770 &	0.4194 &	2.2513	 & 1.4$\times 10^{-7}$ &	49.4932 & 1.6$\times 10^1$ &	0.0001 &	0.0049 &	1.6$\times 10^1$  \\
5.7457 &	0.2906 &	1.8077	 & 9.0$\times 10^{-9}$ &	26.8568 &	3.2 &	0.0001 &	0.0027 &	3.2  \\
4.8839 &	0.1500 &	1.1372	 & 1.5$\times 10^{-8}$ &	13.7707 &	2.3$\times 10^{-1}$ &	0.0001 &	0.0014 &	2.3$\times 10^{-1}$  \\
4.2468 &	0.0001 &	0.0006	 & 6.5$\times 10^{-2}$ &	6.0951  & 6.5$\times 10^{-2}$ &	0.1005 &	1.0170 &	6.0$\times 10^{-7}$  \\
3.9071 &	0.0001 &	0.0003	 & 2.9$\times 10^{-1}$ &	2.8198  &   2.9$\times 10^{-1}$ &	0.1536 &	0.9808 &	4.3$\times 10^{-2}$  \\
3.7568 &	0.0001 &	0.0002	 & 4.0$\times 10^{-1}$ &	1.5649  &   4.0$\times 10^{-1}$ &	0.1903 &	0.8483 &	8.5$\times 10^{-2}$  \\
3.6177 &	0.0001 &	0.0001	 & 5.0$\times 10^{-1}$ &	0.5104  & 5.0$\times 10^{-1}$ &	0.2381 &	0.6756 &	1.3$\times 10^{-1}$  \\
3.4885 &	0.0001 &	-0.0000	 & 5.9$\times 10^{-1}$ &	-0.3766 &	5.9$\times 10^{-1}$ &	0.3019 &	0.4511 &	1.6$\times 10^{-1}$  \\
3.3682 &	0.0001 &	-0.0001	 & 6.5$\times 10^{-1}$ &	-1.1241 & 6.5$\times 10^{-1}$ &	0.3898 &	0.1107 &	1.9$\times 10^{-1}$  \\
3.2559 &	0.0001 &	-0.0002	 & 7.1$\times 10^{-1}$ &	-1.7558 & 7.1$\times 10^{-1}$ &	0.5172 &   -0.3566 &	2.1$\times 10^{-1}$  \\
\hline
& (I$-$) &&& (II$-$) && (III$-$) && \\ 
$L$ & $a$ & $b$ & $\Delta\eta$ & $b$ & $\Delta\eta$ & $a$ & $b$ & $\Delta\eta$ \\
\hline
8.8798 &	0.0106 &	1.4862 &	5.4$\times 10^1$ &	97.3074 & 5.6$\times 10^1$ &	0.0001 &	0.0097 &	5.6$\times 10^1$ \\
6.9770 &	0.0188 &	1.6213 &	1.7$\times 10^1$ &	56.8710 &	1.9$\times 10^1$ &	0.0001 &	0.0057 &	1.9$\times 10^1$ \\
5.7457 &	0.0327 &	1.6434 &	4.2 &	34.2346 &	5.1 &	0.0001 &	0.0034 &	5.1 \\
4.8839 &	0.0572 &	1.7306 &	5.9$\times 10^{-1}$ &	21.1485 &	1.1 & 0.0001 &	0.0021 &	1.1 \\
4.2468 &	0.1022 &	1.9351 &	3.6$\times 10^{-3}$ &	13.4728 & 1.4$\times 10^{-1}$ &	0.0001 &	0.0013 &	1.4$\times 10^{-1}$ \\
3.9071 &	0.1250 &	1.4114 &	8.5$\times 10^{-9}$ &	10.1975 &	2.1$\times 10^{-2}$ &	0.0001 &	0.0010 &	2.1$\times 10^{-2}$ \\
3.7568 &	0.1164 &	1.0874 &	1.6$\times 10^{-9}$ &	8.9426  &   5.8$\times 10^{-3}$ &	0.0001 &	0.0009 &	5.8$\times 10^{-3}$ \\
3.6177 &	0.0909 &	0.7270 &	1.1$\times 10^{-9}$ &	7.8882  &   8.4$\times 10^{-4}$ &	0.0001 &	0.0008 &	8.4$\times 10^{-4}$ \\
3.4885 &	0.0001 &	0.0007 &	2.3$\times 10^{-7}$ &	7.0012  & 2.3$\times 10^{-7}$ &	0.0147 &	0.1029 &	1.4$\times 10^{-9}$ \\
3.3682 &	0.0001 &	0.0006 &	2.6$\times 10^{-4}$ &	6.2537  & 2.6$\times 10^{-4}$ &	0.1089 &	0.6775 &	2.2$\times 10^{-9}$ \\
3.2559 &	0.0001 &	0.0006 &	4.7$\times 10^{-4}$ &	5.6220  &   4.7$\times 10^{-4}$ &	0.1629 &	0.9108 &	3.3$\times 10^{-9}$ \\
\hline
\end{tabular}
\caption{Fit of the solutions of Table \ref{table_conv_series_1} with the six
possible models given by equations \eqref{I},\eqref{II}, and \eqref{III}.}\label{table_fitting_etas} 
}
\end{center}
\end{table}
It is very interesting to see how well the $z$-averaged $\eta$ function of our
numerically computed solution fits, in the whole $\rho$ range, one of the
model $\eta$ functions. As expected, for large values of $L$ the
best fiting model is (III$+$). Then, for decreasing values of $L$ the best
fitting model becomes (I$+$), then to (I$-$) and finally (III$-$).  
Figure \ref{fig_fitting_etas} show six examples taken from Table
\ref{table_fitting_etas}.

The Smarr formula, presented in the usual form, is 
\begin{equation}
M = \frac{1}{4 \pi} \kappa A + 2 \Omega_\Ho J, 
\end{equation}
where $M$ and $J$ are the Komar mass and angular momentum, respectively, $\kappa$ is the horizon temperature, $A$ is the horizon area and $\Omega_\Ho$ is the horizon angular velocity. Since the solutions that reach infinity are those in family (III+), we have that Smarr formula reduces to
\begin{equation}
(1-a)L/4 = \frac{1}{4 \pi} \kappa A + 2 \Omega_\Ho J, 
\end{equation}
where $0 < a \leq 1$. This positivity of $a$ allows us to obtain an interesting bound for $L$, since it can be shown $\Omega_\Ho J > 0$, forcing
\begin{equation} 
0 <2 \Omega_\Ho J \leq \frac{L}{4} - \frac{1}{4 \pi} \kappa A = \frac{L}{4} - m.
\end{equation}
Therefore, there are no solution that extend to infinity whenever $L < 4m$. As we are checking with the numerical analysis, this is a rough lower bound for the critical $L$.

\begin{figure}[t!]
\captionsetup{margin=1cm}
\begin{center}
\includegraphics[width=4.5cm]{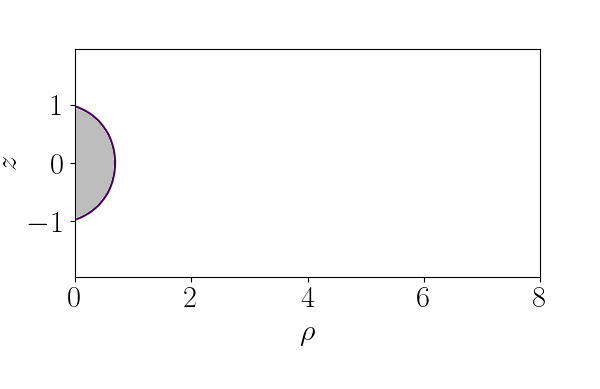}
\includegraphics[width=4.5cm]{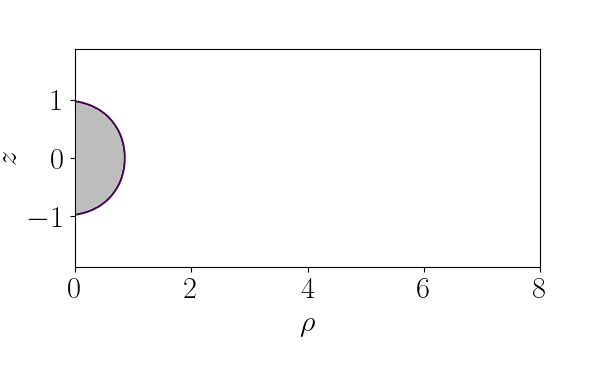}
\includegraphics[width=4.5cm]{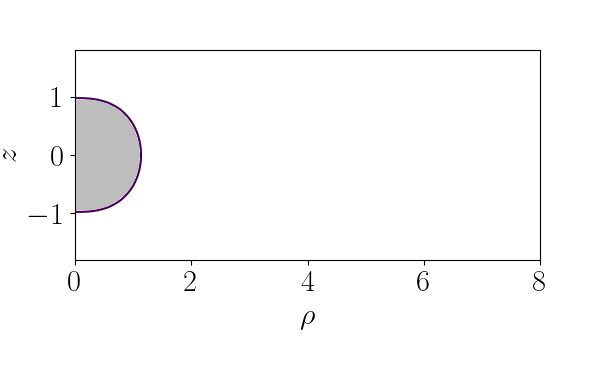}\\
\includegraphics[width=4.5cm]{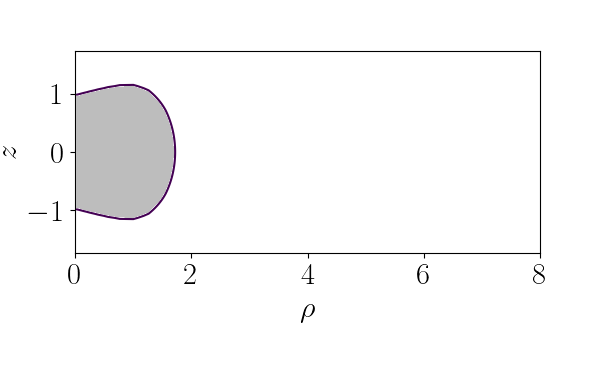}
\includegraphics[width=4.5cm]{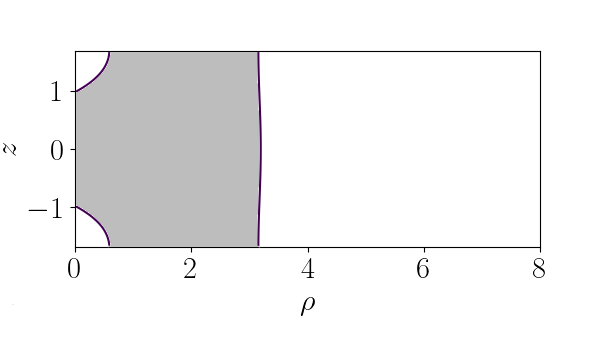}
\includegraphics[width=4.5cm]{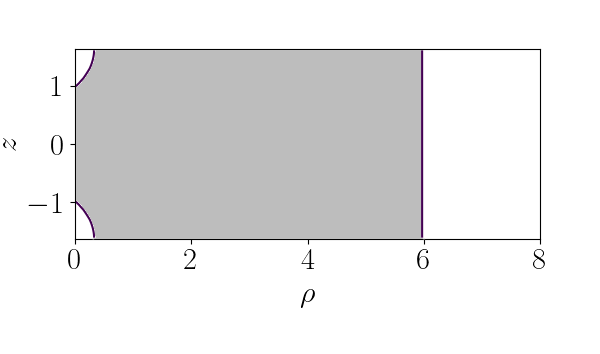}
\caption{The shaded regions are the ergospheres of the last six solutions of
Table \ref{table_conv_series_1}. From left to right, from top to bottom the six
plots correspond $L=8.8798,~ 4.8839,~ 4.2468,~ 3.7568,~ 3.4885,~
3.2559$.}\label{fig_ergosphere_merger}
\end{center}
\end{figure}

\paragraph{Ergosphere merging.} For large values of $L,$ ergospheric region
associated to the horizon does not touch the $z=\pm L/2$ boundaries of the
domain. This ergosphere is topologically $S^2$. When the value of $L$ decreases
the ergospheric region get closer to the boundaries and at some point touches
them. In other words the ergosphere becomes topologically a torus $T^2$. This
change in the topology of the ergosphere is not new, as it has been studied in
binary systems, so is not a surprise to see a development of such
behaviour in the periodic set up. This process is shown in Figure \ref{fig_ergosphere_merger}.

\subsection{Second series: $J=1/2$}
We include here, for the sake of comparison, results corresponding to a series of
four solutions with a higher value of $J$ and the same horizon's area
as the previous series. The three solutions with
larger value of $L$ are better fitted by the asymptotic model (III$+$) of
\eqref{III}, while the solution with $L=4.5475$ is is better fitted with the
model (I$+$) of equation \eqref{I}.

In table \ref{table_results_series_2} the relevant physical quantities of the
solutions in this series are shown. 
\begin{table}[h!]
\captionsetup{margin=1cm}
\begin{center}
{\footnotesize
\begin{tabular}{ccccccc}
$N_h$ & $L$ & $M$ (mass) & Angular velocity &  $\alpha$ (from $M$) & $\alpha$ (from $V$) \\
\hline
22 & 8.2683 & 1.0391 & 1.3024$\times 10^{-1}$ & 5.0272$\times 10^{-1}$ & 5.0278$\times 10^{-1}$ \\
28 & 6.4965 & 1.0518 & 1.4287$\times 10^{-1}$ & 6.4761$\times 10^{-1}$ & 6.4781$\times 10^{-1}$ \\
32 & 5.3501 & 1.0775 & 1.6860$\times 10^{-1}$ & 8.0560$\times 10^{-1}$ & 8.0632$\times 10^{-1}$ \\
40 & 4.5475 & 1.1340 & 2.2516$\times 10^{-1}$ & 9.9745$\times 10^{-1}$ & 1.0003 \\
\hline
\end{tabular}
\caption{Relevant quantities computed for solutions with
$J=1/2$.}\label{table_results_series_2} 
}
\end{center}
\end{table}
\section{Conclusions} \label{conclusions}
In this work we present numerical solutions of the Einstein equations in a periodic set up, for stationary, coaxial and periodic co-rotating black holes, improving the knowledge of stationary solutions outside the standard asymptotically flat scenario and trivial topology. The solutions add angular momentum to the important Myers/Korotkin-Nicolai configurations of static, coaxial and periodic black holes, thus realizing a periodic analogue of the Kerr solution. The solution space contains a rich structure, sharing many qualitative and quantitative properties with the family of Stockum infinite rotating cylinders.
 
Periodic Kerr solutions and Stockum solutions have both Lewis asymptotic. Furthermore, both families posses a critical phenomenon: in the case of periodic Kerr we find that numerical solutions with a given value for the area $A$ and for the angular momentum $J$ of the horizons appear to exist only when the separation between consecutive horizons is larger than a certain critical value that depends only on $A$ and $|J|$. Below the mentioned critical value the rotational energy appears to be too big to sustain a global equilibrium and a singularity shows up at a finite distance from the bulk. For Stockum cylinders, instead, the asymptotic collapse manifest when the angular momentum (per unit of length) reaches a critical value compared to the mass (per unit of length). In both families the collapse is through a transition in the Lewis models of asymptotic.

The Smarr identity proved to be a powerful tool to define an appropriate Neumann-like asymptotic boundary condition for the harmonic map heat flow. This new dynamical Neumann condition can be used in other periodic configurations, for example in periodic arrangements of counter-rotating black holes, situation that we are presently investigating. In this case, the asymptotic model is not that of a rotating rod, but Kasner as the MKN family. The solution space seems to have quite different properties. This in particular poses interesting questions regarding the collapse of the asymptotic and the existence of solutions when the horizons get closer.

This work explores several features which are important in black hole physics in other contexts as well, such as string theory \cite{Sen:1994eb, Khuri:1994gq, Frolov:2002mq, Gibbons:1985ac, Larsen:1999pu, Larsen:1999pp}, supergravity \cite{Myers:1987qx, Park:1995wk} and binary solutions \cite{Manko2008,Manko2011,Manko2017,Astorino:2021boj}.

\section*{Acknowledgements} 
JP and MR would like to thanks M. Khuri, D. Korotkin, H. Nicolai, G. Weinstein and S. Yamada for helpful discussions, and L. Anderson for giving the opportunity to present the work on a Max Planck Institute seminar and insightful comments. 
\noindent
JP is partially supported by CAP PhD scholarship and by CSIC grant C013-347. 
\noindent
MR and JP are partially supported by PEDECIBA.
\noindent
OO acknowledge partial funding by Grants PIP No. 11220080102479 (CONICET,
Argentina) and ``Consolidar'' 33620180100427CB (SeCyT, UNC). OO also wants to
thank the hospitality of Centro de Matemática (Facultad de Ciencias,
Universidad de la República, Uruguay) during a visit in 2022 related to
work on this paper.

\appendix

\section{Simple deduction of Lewis's models}\label{asymptotic_candidates}

Assume then that $\partial_{z}$ is a Killing field so that the metric
coefficient $V, W, \eta$ and $\gamma$ are $z$-independent. This implies that
$\sigma$ is independent of $z$ and it is simple to see that $\omega$ must be
independent of $\rho$. The equations for $\sigma$ and $\omega$ thus become,
\begin{equation}
\partial^2_{\rho} \sigma + \frac{1}{\rho} \partial_{\rho} \sigma = -
\frac{e^{-2 \sigma} (\partial_{z} \omega )^2}{\rho^4},\quad \partial^2_z \omega = 0.
\end{equation}
Hence $\omega = w z + w_1$ with $w\neq 0$ and $w_{1}$ constants. Therefore,
\begin{equation}
\partial^2_{\rho} \sigma + \frac{1}{\rho} \partial_{\rho} \sigma = - \frac{e^{-2
\sigma} w^2}{\rho^4}.
\end{equation}
After making the change of variables $\bar{u} = -\sigma - \ln \rho$ the previous equation implies,
\begin{equation} \label{t2red}
\left(  \rho \bar{u}' \right)^2 = w^2 e^{2 \bar{u} } + C,
\end{equation}
where $'=d/d\rho$. The sign of the constant $C$ will be relevant. Set $C =
\epsilon a^2$, with $a\geq 0$ and $\epsilon = -1,1$. 

Now, letting $x = e^{-\bar{u}}(=\eta/\rho)$ and defining the new variable $\zeta
= \ln \rho/\rho_0$ (for some $\rho_0 > 0$), we arrive at (we use $\partial_\zeta
x = \dot{x}$), $\dot{x}^2 = w^2 + \epsilon a^2 x^2$, to obtain after
$\zeta$-derivation\footnote{$\dot{x}=0$ or $x= \epsilon w/a$ are not solutions
of the original problem and were introduced in the deduction of \ref{t2red}}, 
\begin{equation}
\ddot{x} = \epsilon a^2 x.
\end{equation}
\begin{figure}[h]
\centering
\captionsetup{margin=1cm}
\includegraphics[scale=0.25]{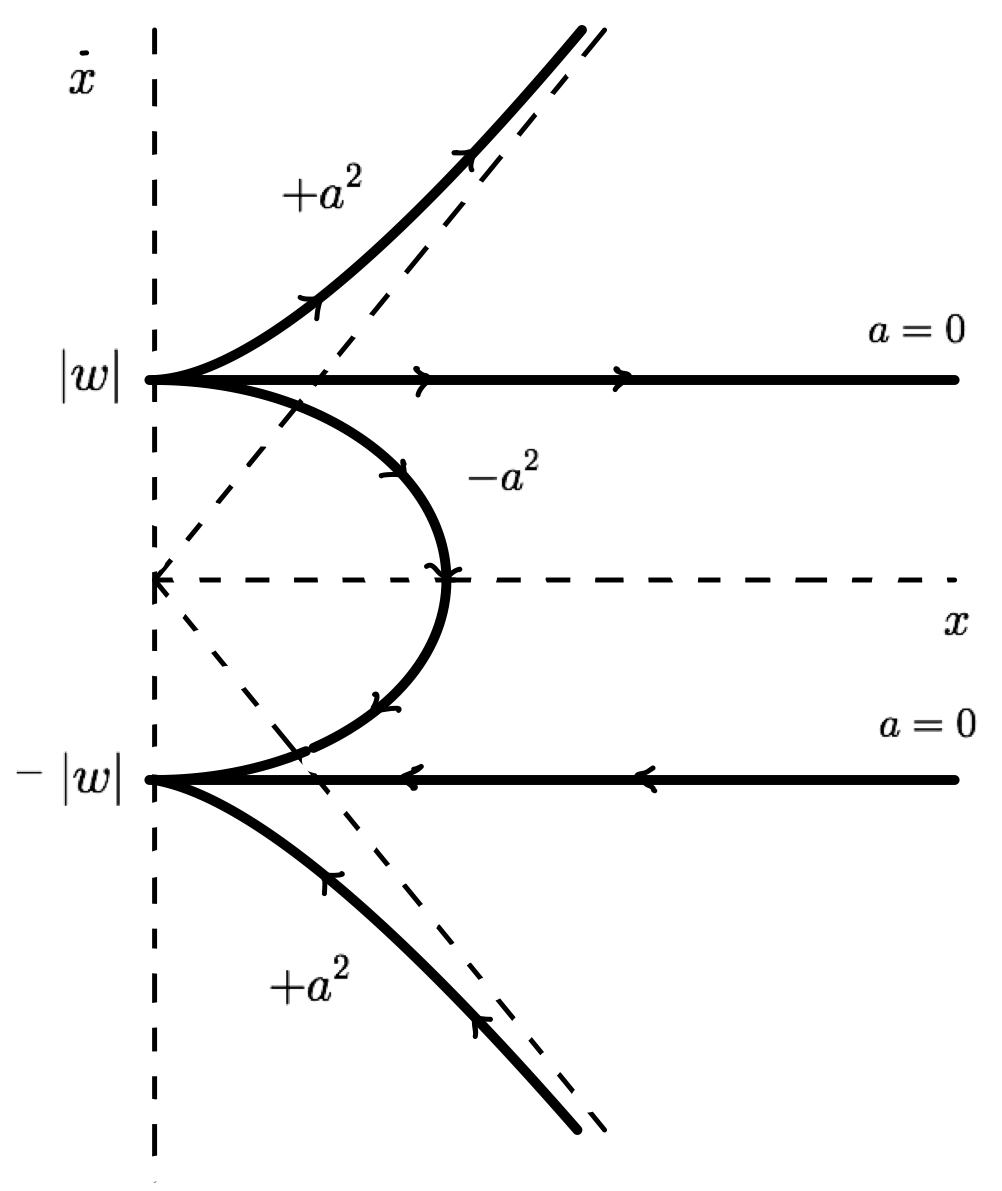}
\caption{Types of $x-$orbits in phase space $(x,\dot{x})$.}\label{orbits}
\end{figure}
%
There are three types of solutions:
\begin{enumerate} 
\item For $a = 0$, we have,
\begin{equation}
x = |w| \left( \pm \ln \rho + b \right),
\end{equation}
\item For $a> 0$ and $\epsilon =-1$, we have,
\begin{equation}\label{asymp_eta_2}
x = \frac{|w|}{a}  \sin ( \pm a \ln \rho + b),
\end{equation}
\item For $a> 0$ and $\epsilon =1$, we have,
\begin{equation} \label{solT2}
x = \frac{|w|}{a} \sinh ( \pm a \ln \rho + b).
\end{equation}
In the three cases $b$ is an arbitrary constant. 

\end{enumerate}

\bibliographystyle{unsrt}
\bibliography{periodicKerrBib.bib}

\end{document}